\documentclass[preprintnumbers, floatfix,letterpaper,aps,prd,epsfig,nofootinbib,
twocolumn
]{revtex4-1}
\pdfoutput=1
\usepackage{bm,graphicx,dcolumn,epstopdf,epsf, latexsym,mathbbol, amssymb,amsmath,color,slashed, mathrsfs,mathcomp, simplewick}
\usepackage[center]{subfigure}
\usepackage{multirow}
\usepackage{makecell}
\usepackage{steinmetz}
\usepackage[colorlinks,linkcolor=blue,citecolor=blue,urlcolor=blue]{hyperref}
\begin{document}
\allowdisplaybreaks
%%%%%%%%%%%%%%%%%%%%%%%%
 \newcommand{\bq}{\begin{equation}}
 \newcommand{\eq}{\end{equation}}
 \newcommand{\bqn}{\begin{eqnarray}}
 \newcommand{\eqn}{\end{eqnarray}}
 \newcommand{\nb}{\nonumber}
 \newcommand{\lb}{\label}
 \newcommand{\f}{\frac}
 \newcommand{\p}{\partial}
\newcommand\degree{^\circ}
%%%%%%%%%%%%%%%%%%%%%%%%%
\newcommand{\PRL}{Phys. Rev. Lett.}
\newcommand{\PLB}{Phys. Lett. B}
\newcommand{\PRD}{Phys. Rev. D}
\newcommand{\CQG}{Class. Quantum Grav.}
\newcommand{\JCAP}{J. Cosmol. Astropart. Phys.}
\newcommand{\JHEP}{J. High. Energy. Phys.}
\newcommand{\red}{\textcolor{black}}
\renewcommand{\theequation}{2.\arabic{equation}} \setcounter{equation}{0}

 %%%%%%%%%%%%%%%%%%%%%%%%

\title{Thin Accretion Disk onto slowly rotating black holes in Einstein-{\AE}ther theory}

\author{Cheng Liu${}^{a, b}$}
\email{liucheng@zjut.edu.cn}

\author{Sen Yang${}^{a, b, c}$}
\email{yangs19@lzu.edu.cn}

\author{Qiang Wu${}^{a, b}$}
\email{wuq@zjut.edu.cn}

\author{Tao Zhu${}^{a, b}$}
\email{zhut05@zjut.edu.cn; Corresponding author}

\affiliation{${}^{a}$Institute for Theoretical Physics \& Cosmology, Zhejiang University of Technology, Hangzhou, 310023, China\\
${}^{b}$ United Center for Gravitational Wave Physics (UCGWP),  Zhejiang University of Technology, Hangzhou, 310023, China  \\
${}^{c}$ Institute of theoretical physics, Lanzhou University,~~Lanzhou 730000, China } 

\date{\today}

\begin{abstract}
The accretion disk is formed by particles moving in closed orbits around a compact object, whose physical properties and the electromagnetic radiation characteristics are determined by the space-time geometry around the compact object. In this paper, we study the physical properties and the optical appearance of the electromagnetic radiation emitted from a thin accretion disk around the two types of the black hole solution in Einstein-\AE ther theory. We investigate in detail the effects of the \ae ther field on the energy flux, temperature distribution, and electromagnetic spectrum of the disk in the two types of slowly rotating Einstein-\AE{}ther black holes. Then we plot the ray-traced redshifted image as well as the intensity and polarization profile of a lensed accretion disk around the two types of Einstein-\AE ther black holes. We found that from the image simulation, the \ae ther field only has a certain effect on the central shadow area of the accretion disk. 
\end{abstract}

\maketitle

\section{Introduction}

Black holes are some of the most extreme objects as a direct inference of the General Relativity (GR). So far, black holes are the strongest gravitational field sources known in our universe, with a pretty high spin and high-intensity magnetic field as usual. These properties make balck holes outstanding laboratories to study both matter and gravity in situations out of reach in terrestrial and astrophysics experiments. Their existence as physical objects has become our consensus due to a series of availabel observations, such as gravitational waves generated due to the merging of black holes observed by the LIGO \& Virgo collaboration experiment \cite{ligo1}, the extraordinary observation of the M87* black hole shadow by the Event Horizon Collaboration \cite{m87,Akiyama:2019brx,Akiyama:2019sww,Akiyama:2019bqs, Akiyama:2019fyp, Akiyama:2019eap}, and the observations of the electromagnetic spectrum emitted from an accretion disk around a black hole \cite{frank, yuan, bambi}. 

% The bright accretion disk surrounding the black hole appears distorted due to the phenomenon of gravitational lensing. The region of the accretion disk behind the black hole also becomes visible due to the bending of light by the black hole. 

An accretion disk is a flattend structure formed by rotating magnetofluid which slowly spirals into a compact central objects. A fraction of the heat, which from the gravitational energy released by the magnetofluid, is converted to radiation which is emitted from the inner part of the accretion disk. When these emitted radiation reaches X-ray, optical or radio astronomy telescopes, it procides the possibility of analyzing its electromagnetic spectrum.  With these observations of black holes in gravitational and electromagnetic spectra, together with their future developments, such as the Next Generation Very Large Array \cite{VLA}, the Thirty Meter Telescope \cite{TMT}, and the BlackHoleCam \cite{BHC}, tests of GR and its alternatives in the strong gravity regime are a hot topic nowadays.

% In the strong gravity regime, the observational aspects of black holes are closely related to a narrow region not far from its event horizon, range from the photon sphere to the accretion disk around the black hole. This region is naturally of great significance, as it is perhaps influenced by the possible higher curvature corrections to the Einstein term in GR.

The standard model of geometrically thin and optically thick accretion disk was first proposed by Shakura and Sunyaev in 1973 \cite{SS}. Then it was developed by Novikov and Thorne \cite{Novikov1973},  and Page and Thorne \cite{Page1974}, and has been successfully applied to astrophysical black hole candidates to explain the features of their observed spectra for many years. 

On the other hand, the experimental verification of symmetry breaking is one of the most powerful tools to understand in which direction to extend the current canon toward more fundamental physical theories. The Lorentz invariance is one of the fundamental principles of GR. However, when one considers the quantization of the gravity, such invariance could be violated in the high-energy regime. In this sense, the Lorentz symmetry can only be treated as an approximate symmetry, which emerges at low energies and is violated at higher energies. With these thoughts, a lot of modified gravity theories have been proposed, such as the Horava-Lifshitz theories of quantum gravity \cite{horava_quantum_2009, Horava, Zhu:2011yu, Wang:2017brl} and Einstein-\AE ther theory \cite{davi, d1, jacobson_einstein-aether_2008,li,bat}.

Einstein-\AE ther gravitational theory is a modification of Einstein’s General Relativity where the kinematic quantities of a unitary time-like vector field, known as aether, is introduced in the gravitational action \cite{davi, d1, jacobson_einstein-aether_2008,li,bat}. One of the important characteristics of the Einstein-\AE ther theory is that it describes the classical limit of Hořava gravity \cite{horava_quantum_2009, Horava}. The presence of the \ae ther field defines a preferred timelike direction that violates the Lorentz symmetry \cite{jacobson_gravity_2001, ding_charged_2015}. The breaking of Lorentz symmetry might occur at the Planck or quantum gravity scales if the spacetime continuum is reduced to a discrete structure, thereby causing spacetime to be an emergent phenomenon. Therefore, it is of great interesting to explore the properties of the accretion disk around black holes with the presence of the \ae ther field. 

Not so long ago, two static, charged, and spherically symmetric black hole solutions have been found in the Einstein-\AE ther theory with two specific combinations of the coupling constants \cite{ding_charged_2015}. Another spherically symmetric black hole solution for a class of coupling constants has also been explored by using numerical calculation, and its analytical representation in the polynomial form has been used in the study of the quasinormal modes in the Einstein-\AE ther theory\cite{Eling:2006ec,Wang2020,Wang2021,Konoplya:2006rv,Konoplya:2006ar}. With these spherically symmetric black hole solutions, one can apply the procedure in Refs.\cite{Wang:2012nv} to generate the slowly rotating black hole solutions, which has been used in the study of the black holes shadow cast in the Einstein-\AE ther theory \cite{Tao_shadow}. These solutions, which contain the effects of the \ae ther field, can be used to investigate the effects of the \ae ther field in the accretion disk and compare with the results of GR.

The aim of our work is to study a geometrically thin and optically thick accretion disk around a charged and slowly rotating black hole in Einstein-\AE ther theory. For an astrophysical black hole, the study of the electromagnetic spectrum from the accretion process around the black hole is a powerful approach to explore the nature of the black hole spacetime in the regime of strong gravity. This has stimulated a lot works on the studies of the thin accretion disk around various black hole spacetimes, see \cite{Mohad,Faraji2020, Harko2009a, Joshi2014, Kovas2010, Zhang:2021hit, Abramowicz2013, Torres, Fish2016, Muller2012, Chow2012, Kovacs2009, Danila, Pun2008, Perez, Gyul2019, Harko2011, Fard2010, Harko2010, Perez2017, Kari2018, Loda, chen2011, chen2012, Liu, spherical, Baner2017, Baner2019,4DEGB} and references therein. Therefore, it is natural to ask whether the \ae ther field corrections of the Einstein-\AE ther theory can appear in the electromagnetic signatures and the optical appearance to a distant observer of the accretion disk. To answer this question, we consider a thin relativistic accretion disk model around the charged and slowly rotating Einstein-\AE ther black hole, which is in a steady-state and in hydrodynamic and thermodynamic equilibrium. In particular, we calculate the energy flux, temperature distribution, and electromagnetic spectrum of the thin accretion disk, and compare them with the standard GR case. Meanwhile, we have analyzed the observable images of the thin accretion dick in its vicinity and the radiation emitted by it, and compared the possible deviation from the case of the Schwarzschild black hole. The possible effects of the \ae ther field corrections on the electromagnetic signatures from the thin accretion disk are also explored.

The content of our paper is as follows. In Sec. II, we present a brief review of the charged and slowly rotating black hole solutions and their properties in the Einstein-\AE ther theory. Then with this rotating solution,  we study the geodesic and the radius of the marginally stable circular orbit vary with the \ae ther field coupling parameter $c_{13}$ and $c_{14}$ in Sec. III. In Sec. IV, we study the properties of the thin accretion disk onto the charged and slowly rotating black hole in Einstein-\AE ther theory and discuss the flux energy of the accretion disk with the effect from \ae ther field. In Sev. V, we present the Ray-traced redshifted image and intensity and polarization profile of a lensed accretion disk around the two classes of black holes solution in Einstein-\AE ther theory considered here, and we comment on the novel features that we obtain by our analysis. The summary and discussion for this paper is presented in Sec. VI.

\section{Black hole solutions in Einstein-\AE ther theory}

In this section, we review briefly the black hole solutions in the Einstein-\AE{}ther theory focusing on its analytic case. \red{Einstein-\AE{}ther theory is essentially general relativity (GR) coupled with a unit, timelike vector field, $u^\mu$. The unit and timelike properties of the \ae{}ther field forces it to be ever-present, even in the local frame, thus selecting a preferred time direction and violating the local Lorentz symmetry. The action for the \AE{}ther theory makes all possible terms that are quadratic in the first derivatives of $u^\mu$. So the Einstein-\AE{}ther theory can be considered as an effective description of Lorentz symmetry breaking in the gravity sector. Otherwise, it has been extensively used in order to obtain quantitative constraints on Lorentz-violating gravity.}

% \subsection{Field Equations of the Einstein-\AE{}ther theory}

%In Einstein-\ae ther theory, in addition to the spacetime metric tensor field $g_{\mu\nu}$, it involves a dynamical, unit timelike \ae ther field $u^{\alpha}$ (it is also called \ae ther four velocity) \cite{jacobson_gravity_2001, foster_radiation_2006, garfinkle_numerical_2007, jacobson_einstein-aether_2008}. Like the metric, and unlike other classical fields, the \ae ther field $u^{\alpha}$ cannot vanish anywhere, so it breaks local Lorentz symmetry. 

\red{With the unit and timelike vector, the general action of the Einstein-\AE{}ther theory is given by \cite{jacobson_einstein-aether_2008}
\bqn
S_{\ae} = \frac{1}{16 \pi G_{\ae}} \int d^4 x \sqrt{-g} \Big(R+ \mathcal{L}_{\ae}\Big),
\eqn
where $g$ is the determinant of the four dimensional metric $g_{\mu\nu}$ of the space-time with the signatures $(-, +, +, +)$, $R$ is the Ricci scalar, $G_{\ae}$ is the \ae{}ther gravitational constant, and the Lagrangiean of the \ae ther field $\mathcal{L}_{\ae}$ is given by
\bqn
\mathcal{L}_{\ae} \equiv -M^{\alpha \beta}_{\;\; \;\; \mu\nu} (D_\alpha u^\mu) (D_{\beta} u^\nu) + \lambda (g_{\mu \nu} u^\mu u^\nu +1).\nb\\
\eqn
Here $D_{\alpha}$ denotes the covariant derivative with respect to $g_{\mu\nu}$, $\lambda$ is a Lagrangian multiplier, which guarantees that the aether four-velocity $u^{\alpha}$ is always timelike, and $M^{\alpha \beta}_{\;\; \;\; \mu\nu}$ is defined as \footnote{The parameters $(c_1, c_2, c_3, c_4)$ used in this paper are related to parameters $(c_\theta, c_\sigma, c_\omega, c_a)$ by the relations $c_{\theta} = c_1+3c_2+c_3$, $c_{\sigma } = c_1+c_3 = c_{13}$, $c_\omega = c_1- c_3$, $c_a = c_1+c_4 = c_{14}$.}
\bqn
M^{\alpha \beta}_{\;\; \;\; \mu\nu} \equiv c_1 g^{\alpha \beta} g_{\mu\nu} + c_2 \delta^{\alpha}_{\mu} \delta ^{\beta}_{\nu} + c_3  \delta^{\alpha}_{\nu} \delta ^{\beta}_{\mu} - c_4 u^{\alpha} u^{\beta} g_{\mu\nu}. \nb\\
\eqn
The four coupling constants $c_i$ ' s are all dimensionless, and $G_{\ae}$ is related to the Newtonian constant $G_N$ via the relation \cite{carroll_lorentz-violating_2004},
\bqn
G_{\ae} = \frac{G_N}{1-\frac{1}{2}c_{14}}
\eqn
with $c_{14} \equiv c_1+c_4$. In order to discuss the black hole solution with electric change, we also add a source-free Maxwell Lagrangian $\mathcal{L}_{M}$ to the theory, then the total action of the theory becomes,
\bqn
S_{\ae, M} =  S_{\ae} + \int d^4x \sqrt{-g } \mathcal{L}_M,
\eqn
where
\bqn
\mathcal{L}_{M} &=& - \frac{1}{16 \pi G_{\ae}} F^{\mu\nu}F_{\mu\nu},\\
F_{\mu\nu} &=& D_\mu A_\nu - D_\nu A_\mu,
\eqn
where $A_\mu$ is the electromagnetic potential four-vector. {It is worth noting that the electromagnetic field $A_\mu$ is minimally coupled to the gravity and the \ae{}ther field $u_\mu$.}
The variations of the total action with respect to $g_{\mu\nu}$, $u^{\alpha}$, $\lambda$, and $A^a$ yield, respectively, the field equations,
\bqn
E^{\mu\nu} =0, \lb{einstein_equation}\\
{\AE}_{\alpha} =0, \lb{aether_equation}\\
g_{\mu\nu} u^{\mu} u^{\nu} =-1 \lb{lambda_equation},\\
D^\mu F_{\mu\nu}=0.
\eqn
where
\bqn
E^{\mu\nu} \equiv R^{\mu\nu} - \frac{1}{2} g^{\mu\nu} R - 8 \pi G_{\ae} T_{\ae}^{\mu\nu},\\
\AE_{\alpha} \equiv D_{\mu} J^{\mu}_{\;\;\;\alpha} + c_4 a_{\mu} D_{\alpha}u^{\mu} + \lambda u_{\alpha},
\eqn
with
\bqn
T^{\ae}_{\alpha\beta} &\equiv& D_{\mu}\Big[J^{\mu}_{\;\;\;(\alpha}u_{\beta)} + J_{(\alpha\beta)}u^{\mu}-u_{(\beta}J_{\alpha)}^{\;\;\;\mu}\Big]\nb\\
&& + c_1\Big[\left(D_{\alpha}u_{\mu}\right)\left(D_{\beta}u^{\mu}\right) - \left(D_{\mu}u_{\alpha}\right)\left(D^{\mu}u_{\beta}\right)\Big]\nb\\
&& + c_4 a_{\alpha}a_{\beta}    + \lambda  u_{\alpha}u_{\beta} - \frac{1}{2}  g_{\alpha\beta} J^{\delta}_{\;\;\sigma} D_{\delta}u^{\sigma},\nb\\
J^{\alpha}_{\;\;\;\mu} &\equiv& M^{\alpha\beta}_{~~~~\mu\nu}D_{\beta}u^{\nu},\nb\\
a^{\mu} &\equiv& u^{\alpha}D_{\alpha}u^{\mu}.
\eqn
From Eqs.(\ref{aether_equation}) and (\ref{lambda_equation}),  we find that
\bqn
\lb{2.7}
\lambda = u_{\beta}D_{\alpha}J^{\alpha\beta} + c_4 a^2,
\eqn
where $a^{2}\equiv a_{\lambda}a^{\lambda}$.
}

\subsection{Static and Charged Spherically Symmetric Einstein-\ae ther Black Holes}

The general form for a static spherically symmetric metric for \red{the Einstein-\AE{}ther black hole spacetimes} can be written in the Eddington-Finklestein coordinate system as
\bqn \lb{metric_EF}
ds^2 = -e(r) dv^2 +2 f(r) dv dr + r^2 (d\theta^2+\sin^2\theta d\phi^2),\nb\\
\eqn
with the corresponding Killing vector $\chi^a$ and the \ae ther vector field $u^{a}$ being given by
\bqn
\chi^a=(1,0,0,0),\;\;\; u^{a}(r) = (\alpha(r), \beta(r), 0, 0 ),
\eqn
where $e(r)$, $f(r)$, $\alpha(r)$, and $\beta(r)$ are functions of $r$ only.
%, and $\gamma_{ij}$ represents the metric of the two-dimensional sphere $S^2$.
The boundary conditions on the metric components are such that the solution is asymptotically flat, while those for the \ae{}ther components are such that
\bqn
\lim_{r \to +\infty} u^a = (1,0,0,0).
\eqn

As shown in \cite{ding_charged_2015}, there exist two types of exact static and charged spherically symmetric black hole solutions in Einstein-\ae ther theory. The first solution corresponds to the special choice of coupling constants $c_{14}=0$ and $c_{123} \neq 0$ where $c_{123} \equiv c_1+c_2+c_3$, while the second solution corresponds to $c_{123}=0$.

\subsubsection{$c_{14}=0$ and $c_{123} \neq 0$}

For the first solution, we have \cite{ding_charged_2015}
\red{ \bqn
e(r) &=& 1- \frac{2M}{r} +\frac{Q^2}{r^2} \nb\\
&&- \frac{27 c_{13}}{256(1-c_{13})} \left(\frac{2M}{r} \right)^4,\lb{e14}\\
f(r) &=& 1,\\
\alpha(r) &=& \Bigg[ \frac{1}{\sqrt{1-c_{13}}} \frac{3\sqrt{3}}{16} \left(\frac{2M}{r}\right)^2 \nb\\
&&~~~  + \sqrt{1-\frac{2M}{r} + \frac{27}{256} \left(\frac{2M}{r}\right)^4}\Bigg]^{-1},\\
\beta(r) &=& - \frac{1}{\sqrt{1-c_{13}}} \frac{3 \sqrt{3}}{16} \left(\frac{2M}{r}\right)^2,
\eqn}
where $M$ and $Q$ are the mass and the electric charge of the black hole spacetime respectively. It is obvious that when $c_{13}=0$, the above solution reduces to the Reissner-Nordstr\"{o}m black hole.

\subsubsection{$c_{123}=0$}

For the second solution, we have \cite{ding_charged_2015}
\red{
\bqn
e(r) &=& 1- \frac{2M}{r} + \frac{Q^2}{(1-c_{13})r^2} \nb\\
&&-\frac{2c_{13} - c_{14}}{8(1-c_{13})} \left(\frac{2M}{r} \right)^2, \lb{e123}\\
f(r) &=& 1,\\
\alpha(r) &=&  \frac{1}{1+\frac{1}{2} \left[\sqrt{\frac{2-c_{14}}{2(1-c_{13})}} -1 \right]\frac{2M}{r}},\\
\beta(r) &=& - \frac{1}{2} \sqrt{\frac{2-c_{14}}{2(1-c_{13})}} \frac{2M}{r}.
\eqn}
In this case, when $c_{13}=0=c_{14}$, it also reduces to the Reissner-Nordstr\"{o}m black hole.

For both solutions, it is convenient to write the metric (\ref{metric_EF}) in the Eddington-Finklestein coordinate system in the form of the usual $(t, \; r,\; \theta, \; \phi)$ coordinates. This can be achieved by using the coordinate transformation
\bqn
dt = dv - \frac{dr }{e(r)}, \;\; dr=dr.
\eqn
Then the metric of the background spacetime turns into the form
\bqn\lb{metric}
ds^2 = - e(r)dt^2+\frac{dr^2}{e(r)} +  r^2 (d\theta^2+\sin^2\theta d\phi^2).
\eqn
In this metric, the \ae{}ther field reads
\bqn
u^a= \left(\alpha(r) - \frac{\beta(r)}{e(r)}, \beta(r), 0, 0\right).
\eqn

%\subsubsection{Killing Horizon}

\subsection{Slowly rotating black holes}

The rotating black hole in the slow rotation limit in general can be described by the well-known Hartle-Thorne metric \cite{hartle_slowly_1968}
\bqn
ds^2 &=& -e(r) dt^2 + \frac{B(r) dr^2}{e(r)} +  r^2 (d\theta^2+\sin^2\theta d\phi^2) \nb\\
&& -\epsilon r^2 \Omega(r, \theta) dt d\phi + \mathcal{O}(\epsilon^2),
\eqn
where $e(r)$ represents the ``seed" static, spherically-symmetric solutions when $\Omega(r, \theta) =0$, $\epsilon$ denotes a small perturbative rotation parameter. For discussions of the black holes in this paper, we consider $B(r)=1$ \cite{Wang:2012nv, barausse_slowly_2016}. The \ae{}ther configuration in the slow-rotation limit is described by \cite{barausse_slowly_2016}
\bqn
u_a dx^a &=&[\beta(r)-e(r) \alpha(r)]dt +\frac{\beta(r)}{e(r)} dr \nb\\
&&+ \epsilon  [\beta(r)-e(r) \alpha(r)]\lambda(r, \theta) \sin^2\theta d \phi + \mathcal{O}(\epsilon^2),\nb\\
\eqn
where $\lambda(r, \theta)$ is related to the \ae{}ther's angular momentum per unit energy by $u_\phi/u_t = \lambda(r, \theta) \sin^2\theta$.

For asymptotically flat boundary condition, as shown in \cite{barausse_slowly_2016}, $\Omega(r,\theta)$ and $\lambda(r, \theta)$ have to be $\theta$-independent, namely $\Omega(r, \theta)=\Omega(r)$ and $\lambda(r, \theta) = \lambda(r)$. Then the \ae{}ther's angular velocity can be written as
\bqn
\psi(r)=\frac{u^\phi}{u^t}= \frac{1}{2}\Omega(r)- \frac{\lambda(r)}{r^2}.
\eqn
The slowly rotating black holes in the Einstein-\AE ther theory have been obtained in \cite{Wang:2012nv, barausse_slowly_2016} and have been discussed in several papers that related to Horava-Lifshitz gravity \cite{Wang:2012nv, Wang:2012nv}. While in these mentioned papers only neutral black holes have been considered, in this subsection, we present the metric of the charged slowly rotating black holes by directly applying the forms in \cite{barausse_slowly_2016} for $c_{14} =0$ but $c_{123} \neq 0$ and $c_{123}=0$ respectively.

\subsubsection{$c_{14} =0$ and $c_{123} \neq 0$}
For this case, there exists slowly rotating black hole solution in the Einstein-\AE ther theory with a spherically symmetric (hypersurface-orthogonal) \ae ther field configuration, which leads to
\bqn
\Omega(r)= \frac{4 J}{r^3}\;\; {\rm and}\;\; \lambda(r)=0.
\eqn
The metric now reads
\bqn\lb{slow1}
ds^2 &=& -e(r) dt^2 + \frac{dr^2}{e(r)} +  r^2 (d\theta^2+\sin^2\theta d\phi^2) \nb\\
&& -\frac{4 M}{r} a \sin^2 \theta dt d\phi + \mathcal{O}(\epsilon^2).  
\eqn
Note that $e(r)$ is given by Eq.~(\ref{e123}).

\subsubsection{$c_{123}=0$}

For this case, there is no closed form for the expression of $\Omega(r)$ and $\lambda(r)$ except in the limit $c_\omega = c_1-c_3 \to \infty$. In \cite{barausse_slowly_2016}, the asymptotic forms for the derivatives $\Omega'(r)$ and $\lambda'(r)$ has been obtained by expanding them for large $r$ and the corresponding integration constants can be determined by using numerical calculation. In the limit $c_\omega \to \infty$, the frame dragging potential $\Omega(r)$ has the form \cite{barausse_slowly_2016}
\bqn
\Omega(r)=\frac{4 J}{4}.
\eqn
Then the metric in this case takes the same form as (\ref{slow1}) but with $e(r)$ given by Eq.~(\ref{e14}). As pointed out in \cite{Wang:2012nv},
the Einstein-\ae{}ther theory in the limit $c_\omega \to \infty$ coincides to the non-projectable Horava-Lifshitz theory of gravity in the infrared limit, thus the solutions in the Einstein-\ae{}ther theory with $c_\omega \to \infty$ are also solutions of the Horava-Lifshitz theory of gravity.

\red{
\subsection{Numerical black hole solution}
Except the black hole solutions in the analytical form, the numerical black hole solution has also been explored in the Einstein-\AE{}ther theory for a class of coupling constants \cite{Eling:2006ec}. For spherically symmetric solution, $c_4$ can be absorbed into other coupling constants. In \cite{Eling:2006ec},  the so-called non-reduced Einstein-\AE{}ther theory is considered, for which $c_3=0$, and then one can use the field redefinition that fixes the coefficient $c_2$ \cite{Eling:2006ec, Konoplya:2006rv, Konoplya:2006ar} ,
\bqn
c_2= - \frac{-2 c_1^3}{2-4 c_1 +3 c_1^3},
\eqn
so that $c_1$ is the only free parameter in the numerical black hole solution in \cite{Eling:2006ec}. With these set-up, the metric for a spherically symmetric static black hole can still be described by (\ref{metric_EF}), in which the functions $e(r)$, $f(r)$, $\alpha(r)$, and $\beta(r)$ are determined by numerical calculations. An approximate form of $e(r)$ and $f(r)$ are given in the polynomial form in \cite{Konoplya:2006rv, Konoplya:2006ar} as
\bqn
e(r) = \frac{\sum_{i=0}^{N_e} a_{i}^{(e)} r^i} {1+\sum_{i=0}^{N_e} b_{i}^{(e)} r^i},\\
f(r)=  \frac{\sum_{i=0}^{N_f} a_{i}^{(f)} r^i} {1+\sum_{i=0}^{N_f} b_{i}^{(f)} r^i},
\eqn
where the coefficients $(a_i^{(e)}, b_i^{(e)}; a_i^{(f)}, b_i^{(f)})$ can be determined by fitting the above analytical form with numerical solutions, which is out of the scope of the current paper. The general features of the this numerical black hole has been discussed in \cite{Eling:2006ec, Konoplya:2006rv, Konoplya:2006ar}, and it has been shown that the \ae{}ther parameter $c_1$ trends to decrease the radius of the horizon. }

\red{ In addition, Ref \cite{Wang2020} also gives other various numerical solutions of the Einstein-\AE{}ther theory more recently, which with an accuracy that is at least two orders higher than previous ones. We will not introduce and discuss their numerical solutions in detail here.
}

\section{The Geodesic of slowly rotating black holes in Einstein-{\AE}ther theory}
The accretion disk is formed by particles moving in circular orbits around a compact object, whose physical properties and the electromagnetic radiation characteristics are determined by the space-time geometry around the compact object. For the purpose to study the electromagnetic properties of the thin accretion disk around a charged and slowly rotating black hole in Einstein-\AE ther theory, let us first consider the evolution of a massive particle in the black hole spacetime. We start with the Lagrangian of the particle,
\bqn
L = \frac{1}{2}g_{\mu \nu} \frac{d x^\mu} {d \lambda } \frac{d x^\nu}{d \lambda},
\eqn
where $\lambda$ denotes the affine parameter of the world line of the particle. For massless particle we have $L=0$ and for massive one $L <0$. Then the generalized momentum $p_\mu$ of the particle can be obtained via
\bqn
p_{\mu} = \frac{\partial L}{\partial \dot x^{\mu}} = g_{\mu\nu} \dot x^\nu,
\eqn
which leads to four equations of motions for a particle with energy $\tilde{E}$ and angular momentum $\tilde{L}$,
\bqn
p_t &=& g_{tt} \dot t +g_{t\phi} \dot \phi  = - \tilde{E},\\
p_\phi &=&g_{\phi t} \dot t+ g_{\phi \phi} \dot \phi = \tilde{L}, \\
p_r &=& g_{rr} \dot r,\\
p_\theta &=& g_{\theta \theta} \dot \theta.
\eqn
Here a dot denotes the derivative with respect to the affine parameter $\lambda$ of the geodesics. From these expressions we obtain 
\bqn
\dot t =  \frac{g_{\phi\phi} \tilde{E}+g_{t\phi}\tilde{L}  }{ g_{t\phi}g_{\phi t}- g_{tt}g_{\phi\phi} } =\frac{r^4 \tilde{E}-2Mra\tilde{L}}{4M^2a^2\sin^2\theta+r^2e(r)},~~~~~~~\\
\dot \phi = \frac{\tilde{E}g_{t\phi}+ g_{tt} \tilde{L}}{g_{tt}g_{\phi\phi}-g_{t\phi}g_{\phi t}} = \frac{2Mar \sin^2\theta\tilde{E}+ r^2 e(r)\tilde{L}}{r^4e(r)\sin^2\theta+4M^2a^2 \sin^4\theta}.\nb\\
\eqn
Note that in the derivation of the above equation we have used the metric in (\ref{slow1}). For timelike geodesics, we have $ g_{\mu \nu} \dot x^\mu \dot x^\nu = -1$. Substituting $\dot t$ and $\dot \phi$ we can get
\bqn
g_{rr} \dot r^2 + g_{\theta \theta} \dot \theta^2 = -1 - g_{tt} \dot t^2  - g_{\phi\phi}\dot \phi^2 -2g_{t \phi}\dot{t}\dot{\phi}~~~~ \nb\\
= -1 +\frac{r(r^3 \tilde{E}^2+4aM\tilde{E}\tilde{L}+r e(r)\tilde{L}^2\csc^2\theta)}{4a^2M^2\sin^2\theta-r^4e(r)}.
\eqn

We are interested in the evolution of the particle in the equatorial circular orbits. For this reason, we can consider $\theta=\pi/2$ and $\dot \theta=0$ for simplicity. Then the above expression can be simplified into the form
\bqn
\dot r ^2 = V_{\text{eff}}(r,M,\tilde{E},\tilde{L}),                                     
\eqn
where $V_{\rm eff}(r)$ denotes the effective potential of the test particle with energy $\tilde{E}$ and axial component of the angular momentum $\tilde{L}$, which is given by

\bqn \lb{Veff}
V_{\rm eff}(r)= \frac{1}{e(r)}-\frac{r^4\tilde{E}^2+4arM\tilde{E}\tilde{L}+r^2 e(r)\tilde{L}^2}{4a^2M^2e(r)-r^4 e^2(r)}.
\eqn

If we take the $e(r)$ defined by (\ref{e123}), one immediately observes that $V_{\rm eff}(r) \to 1$  as $r \to +\infty$, as expected for an asymptotically flat spacetime. With this case, the particles with energy  $\tilde{E} >1$ can escape to infinity, and $\tilde{E} = 1$ is the critical case between bound and unbound orbits. In this sense, the maximum energy for the bound orbits is $\tilde{E}=1$. Physically, the emergence of the coupling constant has a correction to the structure of spacetime, resulting in a change in the effective potential, naturally, as we'll discuss next, it also affects the size of the radius of the marginally stable circular orbit. The stable circular orbits in the equatorial plane are corresponding to those orbits with constant $r$, i.e., $\dot r^2=0$ and $dV_{\rm eff}(r)/dr=0$. With these conditions, one can write the specific energy $\tilde{E}$, the specific angular momentum $\tilde{L}$, and the angular velocity $\Omega$ of the particle moving in a circular orbit in the black hole as 
\bqn
\tilde{E}&=&-\frac{g_{tt}+g_{t\phi}\Omega}{\sqrt{-g_{tt}-2g_{t\phi}\Omega-g_{\phi\phi}\Omega^2}} , \lb{Etilde}\\
\tilde{L}&=&\frac{g_{r\phi}+g_{\phi\phi}\Omega}{\sqrt{-g_{tt}-2g_{t\phi}\Omega-g_{\phi\phi}\Omega^2}}.   \lb{ltilde}
\eqn
Since the test particles follow geodesic, equatorial, and circular orbits. We can write the geodesic equations as
\bqn
\frac{d}{d\lambda}(g_{\mu\nu}\dot x^\nu)-\frac{1}{2} (\partial_\mu g_{\nu\rho})\dot x^\nu \dot x^\rho=0,   \label{geo}
\eqn
with the conditions $\dot r=\dot \theta= \ddot{r}=0$ for equatorial circular orbits, the radial component of Eq.(\ref{geo}) reduces to 
\bqn
(\partial_r g_{tt})\dot t^2 +2(\partial_r g_{t\phi}) \dot t \dot \phi +(\partial_r g_{\phi\phi})\dot \phi^2=0.
\eqn
Therefore, we get the angular velocity 
\bqn
\Omega=\frac{d\phi}{dt}=\frac{-\partial_r g_{t\phi}\pm \sqrt{(\partial_r g_{t\phi})^2-(\partial_r g_{tt})(\partial_r g_{\phi\phi})}}{\partial_r g_{\phi\phi}}.~~~
\eqn

The marginally stable circular orbits around the slowly rotating Einstein-\AE ther black hole can be determined from the condition 
\bqn
d^2V_{\rm eff}(r)/dr^2 =0.
\eqn
\begin{figure*}
\includegraphics[width=8.1cm]{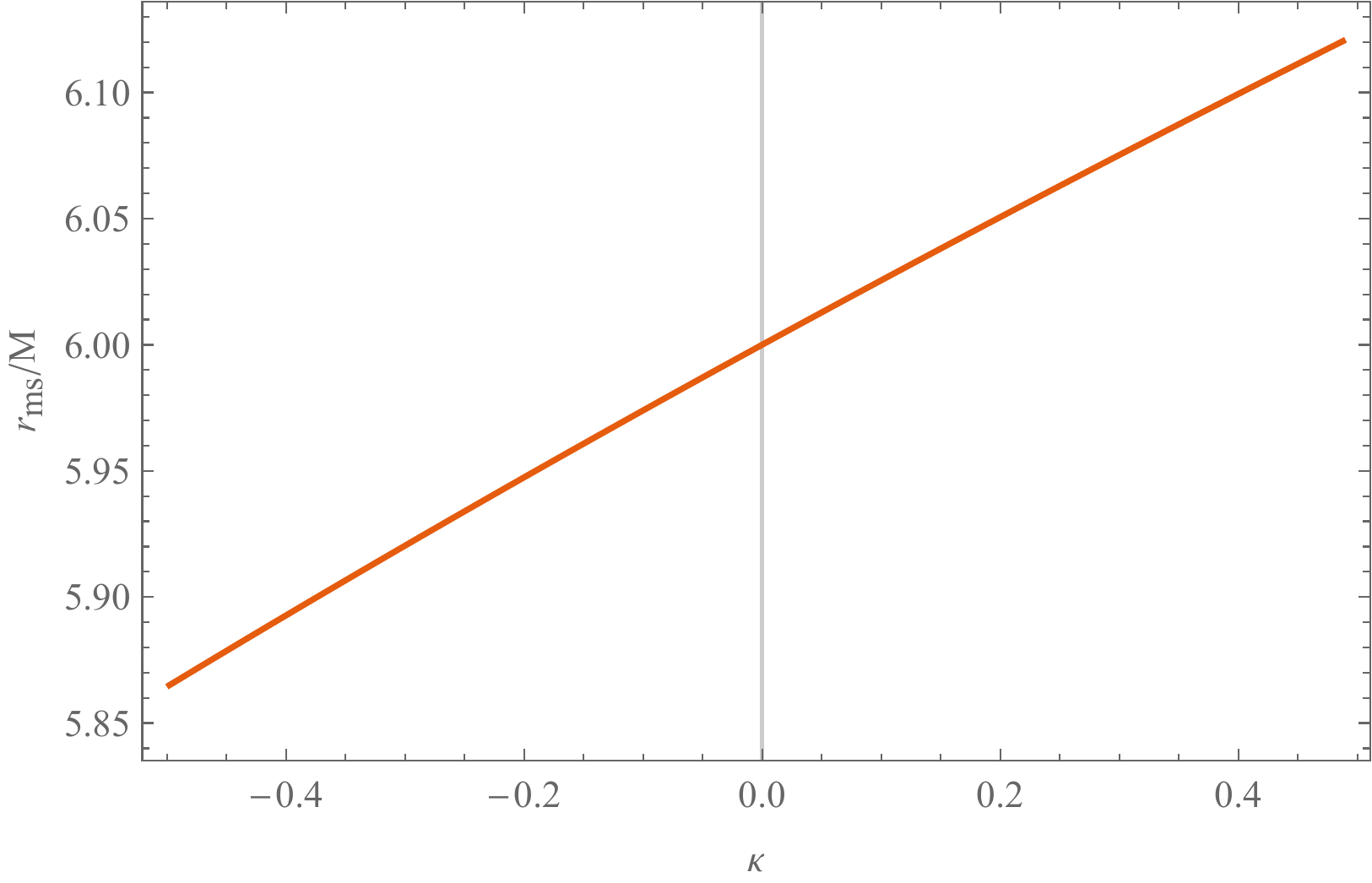}
\includegraphics[width=8.1cm]{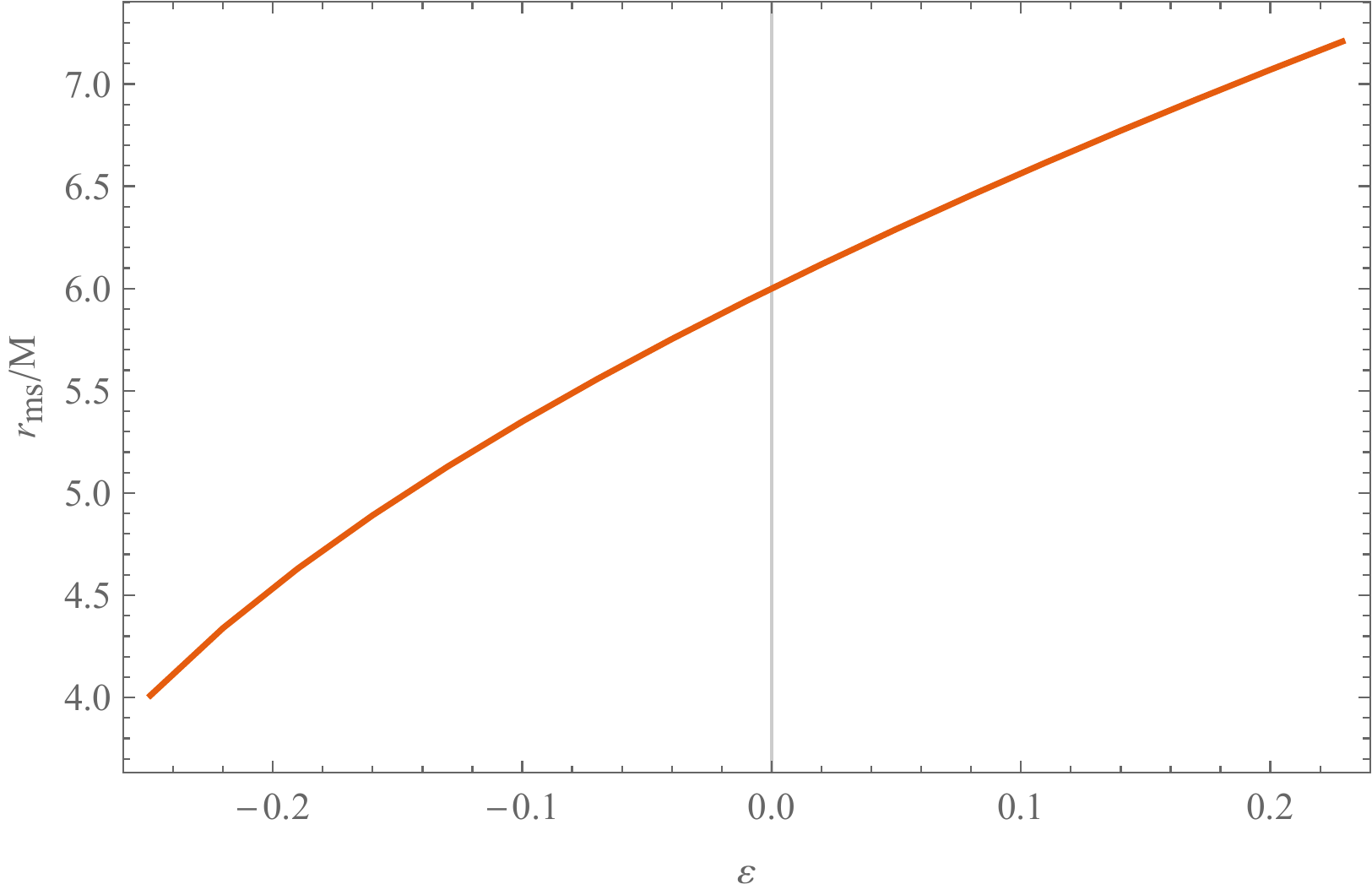}
\caption{\red{Impact of the \ae ther coupling constant $\kappa$ (left panel) and $\varepsilon$ (right panel) on the marginally stable orbit radius $r_{\text{ms}}$ for particles moving in the thin accretion disk around the first and second Einstein-\AE ther black hole, respectively. Here we only considered the neutral black hole case.}} \label{rms}
\end{figure*}
\red{
In order to conveniently characterize the influence of the \ae{}ther field on the space-time around the black hole, we redefine the following two quantities
\bqn
\kappa \equiv \frac{c_{13}}{1-c_{13}}, ~~~ \varepsilon\equiv \frac{2c_{13}-c_{14}}{8(1-c_{13})}.
\eqn
They correspond to the first type solution and the second type solution in the Einstein-\ae{}ther theory, respectively.}
Combining this equation with (\ref{Etilde}) and (\ref{ltilde}) and solving for $r$, the radius of the marginally stable circular orbit can be calculated numerically, which does not admit any analytical solution. We solve it numerically and plot the result in Fig.~\ref{rms}, which shows clearly the radius of the marginally stable circular orbit $r_{\rm ms}$ is increasing as the increasing of the \ae ther field coupling constant $\kappa$ for the first type Einstein-\AE{}ther black hole and $\varepsilon$ for the second type. 

\red{ In addition, we give an approximate analytic result of the marginally stable circular orbits for the first type Einstein-\AE{}ther black hole by making Taylor expansion when $\kappa$ is small with $a=0$, which is
\bqn
r_{\text{ms}}\simeq 6 M\left(1+\frac{11 M \kappa }{256}\right).\lb{rms-c13}
%-Q^2 \left(\frac{3}{2 M}-\frac{25 \kappa }{128 M}\right). 
\eqn
\red{Similarly, we can give an approximate analytic result for the second type solution when $c_{13}$ and $ c_{14} $ are small with $a=0$, that is
\bqn
r_{\text{ms}}\simeq 6 M(1+\varepsilon).\lb{rms_epsilon}
%+ Q^2 \left[\frac{19 \varepsilon}{9 M}-\frac{3}{2M} (1+\kappa) \right].~~ \lb{rms-c14}
% 6 M+c_{13} \frac{3 M}{2}-c_{14}\frac{3  M}{4} ~~~~~~~~~~~~~~\nb\\  -Q^2 \left( \frac{3}{2 M}+c_{13}\frac{35}{36 M} +c_{14}\frac{19 }{72 M} \right).
\eqn}}
These approximate analytical results clearly show a key difference between the charged and slowly rotating Einstein-\AE ther black hole and the Schwarzschild black hole in general relativity, which is also an important physical quantity that is not difficult to observe.

\section{The Properties of Thin Accretion Disk onto slowly rotating black holes in Einstein-{\AE}ther theory}
In this section we will apply the steady-state thin accretion disk model to study the accretion process around \red{the slowly rotating black hole} in Einstein-\AE ther theory. \red{As we all know, the usual astrophysical black holes are neutral, so we also choose to discuss zero charged case here.} Novikov, Thorne, and Page \cite{Novikov1973,Page1974} made a relativistic analysis of the structure of an accretion disk around a black hole. They assumed that the background space-time geometry was static, axisymmetric, asymptotically flat, and reflectively symmetric. They also assumed that the central plane of the disk overlapped with the equatorial plane of the black hole. This assumption leads to measure coefficient of $g_ {tt} $, $g_ {t\phi } $, $g_ {rr} $, $g_ {\theta \theta}$ and $g_ {\phi \phi}$ only depends on the radial coordinates $r$. In this model, the stress-energy tensor of the accreting matter in the disk is decomposed according to
\bqn
T^{\mu\nu}=\rho_0 u^\mu u^\nu +2 u^{(\mu}q^{\nu)}+t^{\mu\nu},
\eqn
where
\bqn
u_\mu q^\mu=0, ~~~u_\mu t^{\mu\nu}=0
\eqn
where the quantities $\rho_0$, $q^\mu$ and $t^{\mu\nu}$ represents the rest mass density, the energy flow vector and the stress tensor of the accreting matter, respectively, which is defined in the averaged rest-frame of the orbiting particle with four-velocity $u^\mu$. From the equation if the rest mass conservation, $\nabla_\mu (\rho_0 u^\mu)=0$, it follows that the time averaged rate of the accretion of the rest mass is independent of the disk radius,
\bqn
\dot{M}_0 &\equiv& -2\pi \sqrt{-g} \Sigma u^r=\text{constant},~~~\nb\\
\Sigma (r)&=&\int^H_{-H}<\rho_0>dz.
\eqn
where $\Sigma(r)$ is the averaged rest mass density. According to the conservation law of the energy and the law of the angular momentum conservation
\bqn
\nabla_\mu(\rho_0 u^\mu)=0, ~~~\nabla_\mu J^\mu=0,
\eqn
we have the integral form
\bqn
&&[\dot{M}_0\tilde{E}-2\pi \sqrt{-g} \Omega W^r_\phi]_{,r} =4 \pi r F(r)\tilde{E}, \\
&&~~[\dot{M}_0\tilde{L}-2\pi \sqrt{-g} W^r_\phi]_{,r}=4 \pi r F(r) \tilde{L}.
\eqn
with\red{
\bqn
W^r_\phi=\int^H_{-H}<t^r_\phi>dz, %~~~\sqrt{-G}= \sqrt{1+L}r.
\eqn}
where $W^r_\phi$ is the averaged torque. The quantity $<t^r_\phi>$ is the average value of the $\phi$-$r$ component of the stress tensor over a characteristic time scale $\Delta t$ and the azimuthal angle $\Delta \phi=2 \pi$. By applying the energy-angular momentum relation $\tilde{E}_{,r}=\omega \tilde{L}_{,r}$, the flux $F(r)$ of the radiant energy over the disk can be expressed in terms of the specific energy, angular momentum, and of the angular velocity of the black hole,
\bqn 
F(r)=-\frac{\dot{M}_0}{4\pi \sqrt{-g}}\frac{\Omega_{,r}}{(\tilde{E}-\Omega\tilde{L})^2} \int^r_{r_{\text{ms}}}(\tilde{E} -\Omega \tilde{L})\tilde{L}_{,r} dr,~~~~~\lb{energyflux}
\eqn
where $r_{\text{ms}}$ is the inner edge of the thin accretion disk and is assumed to be at the radius of the marginally stable circular orbit around the slowly rotating black holes in Einstein-\AE ther theory. The small rotation parameter $a=J/M \ll1$ can only affect the radius and the radiation $F(r)$ of the thin accretion disk if we consider its effects beyond the leading order. In this case, one can still use Eq.(\ref{rms-c13}) and (\ref{rms_epsilon}) to determine the radiation flux of the slowly rotating black holes.

\begin{figure*}
\centering
\includegraphics[width=3.4in]{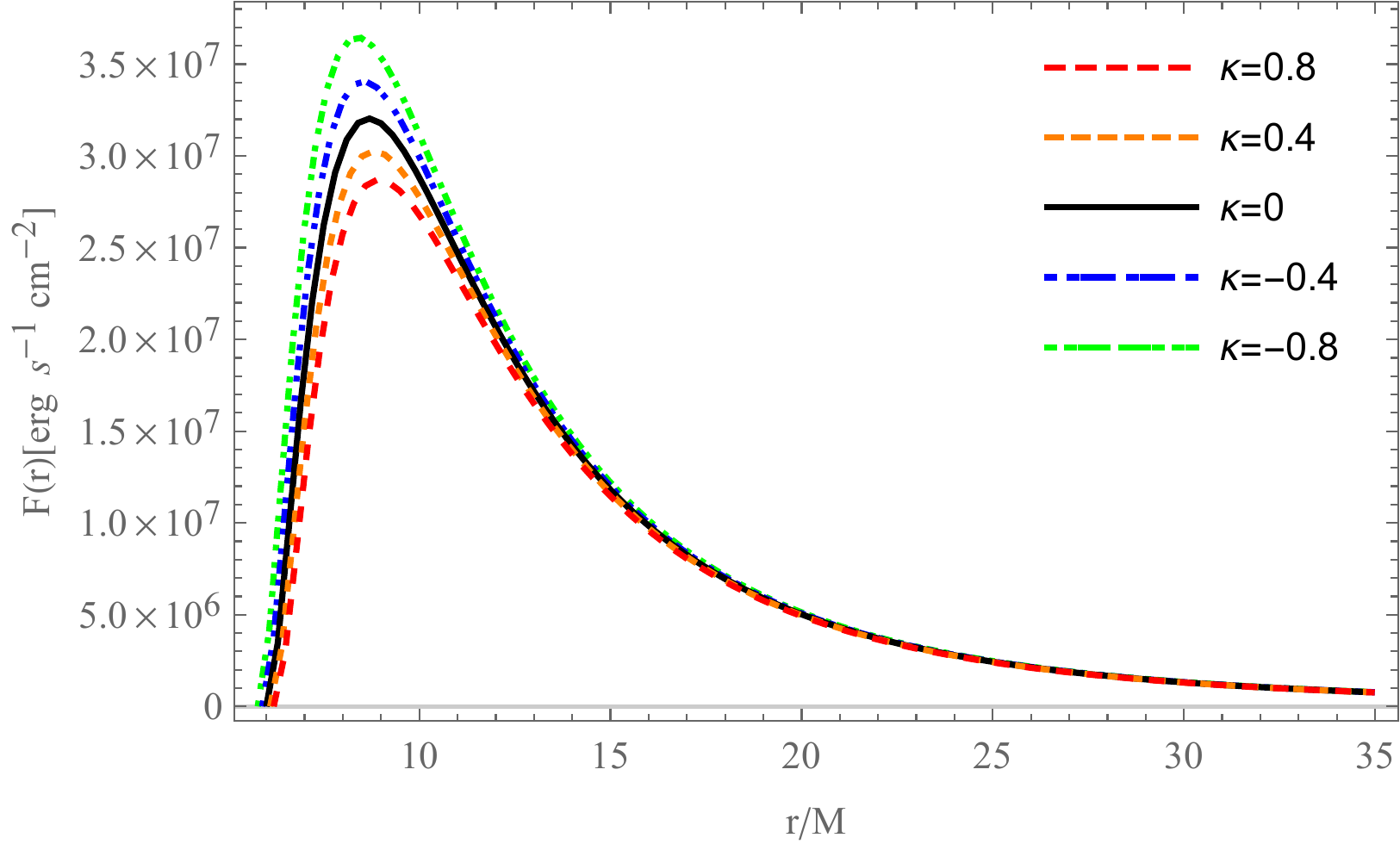}
\includegraphics[width=3.4in]{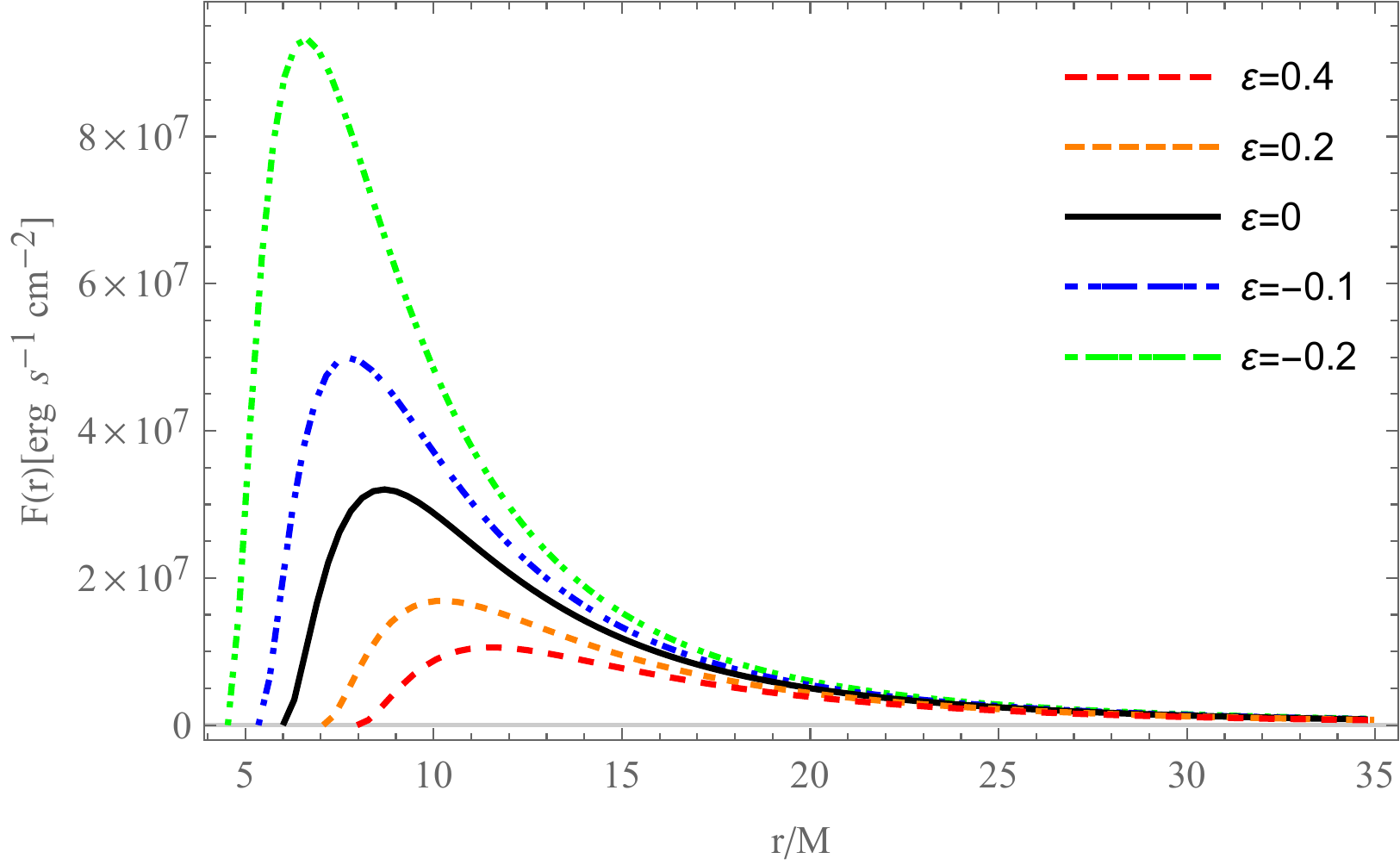}
\caption{\red{Dependence of the radiated energy flux over the thin accretion disk on the radial distance for different values of the \ae ther field coupling constant $\kappa$(left panel) for the first solution and $\varepsilon$(right panel) for the second one. Here the mass of the black hole and the mass accretion rate are set to be $10^{6} M_{\odot}$ and $10^{-12}M_{\odot}/{\rm yr}$, respectively. And we only considered $Q=0$, the neutral black hole case.}}
\label{flux}
\end{figure*}

\red{We calculate the radiation flux $F(r)$ numerically and illustrate its behavior as a function of the radial distance for different solutions of the Einstein-\AE{}ther theory with the \ae ther field coupling constants $\kappa$ and $\varepsilon$. Following \cite{chen2011, chen2012}, we here consider the mass accretion driven by both the Einstein-\AE ther black holes type I and type II with a total mass $M=10^6 M_{\odot}$ with a mass accretion rate of $\dot M_0 = 10^{-12} M_{\odot} /{\rm yr}$. In Fig.~\ref{flux}, we present that the energy flux profile $F(r)$ radiated by a thin accretion disk around the different solutions of the Einstein-\AE{}ther theory with different \ae ther field coupling constants $\kappa$ and $\varepsilon$. For the first solution, It is shown that the energy flux grows monotonically with decreasing the value of \ae ther field coupling constant $\kappa$. In addition, one observes that the energy flux possesses a single maximum, which grows also monotonically with the decreasing of the value of coupling constant $\kappa$. At the same time its radial position is shifted towards the location of the horizon. The main reason is that for negative $\kappa$, the effect of the \ae ther field coupling constant $\kappa$ decreases the radius of the marginally stable orbit so that the lower limit of the integral in (\ref{energyflux}) becomes smaller, while for positive $\kappa$ the radius of the marginally stable orbit increases so that the lower limit becomes larger. 
For the second type of black hole solution, we can also see the same variation law of energy flux. Comparing further with the solutions of the first type, one can easily find that the \ae{}ther field coupling constant $\varepsilon$ variation in the case of the second type of solution has a much greater impact on the energy flux than the first type of solution, although in comparison the second type of solution the coupling constant value range of the solution is much smaller. The reason for such an obvious difference is that the \ae{}ther field correction of the first type of black hole solution is on the fourth-order term of $M/r$, while the \ae{}ther field correction of the second type of black hole solution is on the second-order term.}

\begin{figure*}
\centering
\includegraphics[width=3.4in]{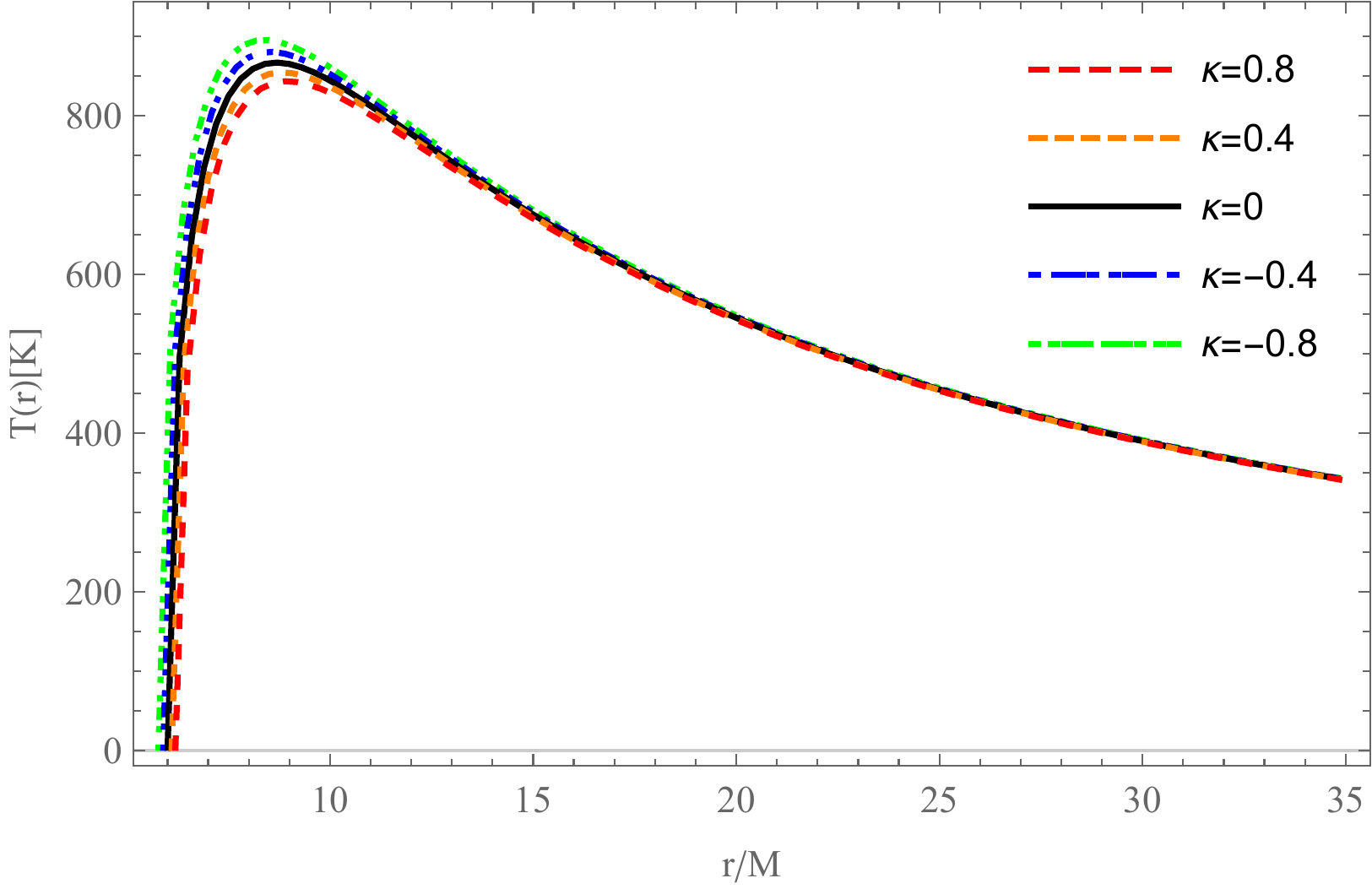}
\includegraphics[width=3.4in]{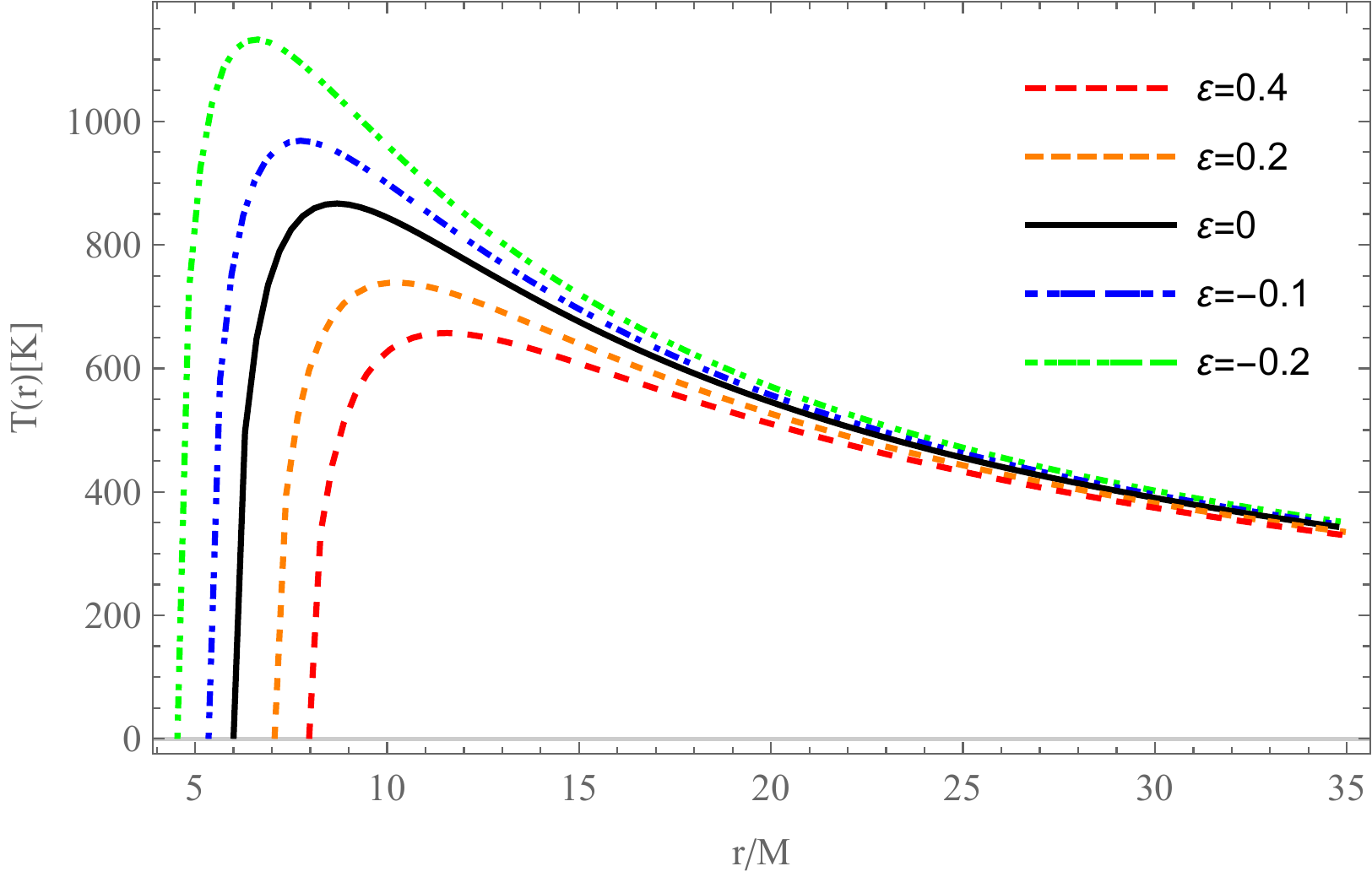}
\caption{\red{The temperature profile of the thin accretion disk around the Einstein-\AE ther black hole for different values of the \ae ther field coupling constant $\kappa$(left panel) for the first solution and $\varepsilon$(right panel) for the second one. Here the mass of the black hole and the mass accretion rate are set to be $10^{6} M_{\odot}$ and $10^{-12}M_{\odot}/{\rm yr}$, respectively. And we only considered $Q=0$, the neutral black hole case.}}
\label{Teff}
\end{figure*}

The accreting matter in the steady state thin disk model is supposed to be in thermodynamic equilibrium. The radiation flux $F(r)$ emitted by the thin accretion disk surface will follow Stefan-Boltzmann law. Therefore, the effective temperature of a geometrically thin black-body disk is given by
\bqn
T_{\rm eff}(r) = \left(\frac{F(r)}{\sigma}\right)^{1/4},
\eqn
where $\sigma = 5.67 \times 10^{-5}\; {\rm erg}\; s^{-1} \;{\rm cm}^{-2}\; K^{-4}$ is the Stefan-Boltzmann constant. 
\red{In Fig.~\ref{Teff}, we display the radial profile of the effective temperature $T_{\rm eff}(r)$ of the thin accretion disk around the different type of black hole solutions with different \ae ther field coupling constants $\kappa$ and $\varepsilon$. The figure of the effective temperature shows a similar behavior as that of the energy flux in Fig.~\ref{flux}. It is easy to see from Fig.~\ref{Teff} that the temperature at the fixed radius grows monotonically with the decreasing the value of $\kappa$ and $\varepsilon$. For a positive value of the \ae ther field coupling constant $\kappa$ and $\varepsilon$, the accretion disk is colder than that around a Schwarzschild black hole. Similarly, the variation of the effective temperature of the thin accretion disk of the second type of black hole solution with the \ae{}ther field coupling constant is much more obvious than that of the first type of solution. This is also caused by the different order of the correction term of the \ae{}ther field.} 
%Similarly, we can see that the existence of the black hole charges grows the temperature of the thin accretion disk around the Einstein-\AE{}ther black holes.

\red{Since we consider the radiation emitted by the thin accretion disk surface as a perfect black body radiation, the observed luminosity $L(\nu)$ of the thin accretion disk around the Einstein-\AE{}ther black hole type I and type II has a red-shifted black body spectrum \cite{Cosimo1, Cosimo2},}
\bqn
L(\nu) &=& 4 \pi d^2 I(\nu) \nb\\
&=& \frac{8 \pi h \cos i}{c^2} \int_{r_i}^{r_f} \int_{0}^{2\pi} g^3 \frac{\nu_e^3 r d \phi dr}{\exp{(\frac{h \nu_e}{k_{\rm B} T})}-1},
\eqn
where $i$ is the inclination angle of the thin accretion disk around the Einstein-\AE{}ther spacetime, $d$ is the distance between the observer and the center of the thin accretion disk, $r_{i}$ and $r_f$ are the inner and outer radii of the disc, $h$ is the Planck constant, $\nu_e$ is the emission frequency in the local rest frame of the emitter, $I(\nu)$ is the Planck distribution, $k_{\rm B}$ is the Boltzmann constant, and $g$ is the redshift factor
\bqn
g=\frac{\nu}{\nu_e}=\frac{k_\mu u^\mu_o}{k_\mu u^\mu_e},
\eqn
where $\nu$ is the radiation fraquency in the local rest frame of the distant observer, $u^\mu_o=(1,0,0,0)$ is the 4-velosity of the observer, and $u^\mu_e=(u^t_e,0,0,\Omega u^t_e)$ is the 4-velosity of the emitter. \red{With the normalization condition, we can write the $u^t_e$ as 
\bqn
u^t_e= \frac{1}{\sqrt{-g_{tt}-2g_{t\phi}\Omega -g_{\phi\phi}\Omega^2}}.
\eqn
According to the $t-$ and $\phi-$component of a photon’s four momentum are conserved quantities in any static spherically symmetric spacetime, we can obtain the quantity
\bqn
\frac{k_\phi}{k_t} = r \sin i \sin \phi,
\eqn
then the redshift factor can be computed to be
\bqn
g=\frac{\sqrt{-g_{tt}-2g_{t\phi}\Omega -g_{\phi\phi}\Omega^2}}{1+\Omega r \sin i \sin \phi}.
\eqn
This way, we can take the relativistic effects of Doppler boost, gravitational redshift, and frame dragging into account with $g$.} Since the flux over the disk surface vanishes at $r \to +\infty$ for asymptotically flat geometry, in this paper, we can take $r_i=r_{\rm ms}$ and $r_f = +\infty$. 
\begin{figure*} 
\includegraphics[width=3.4in]{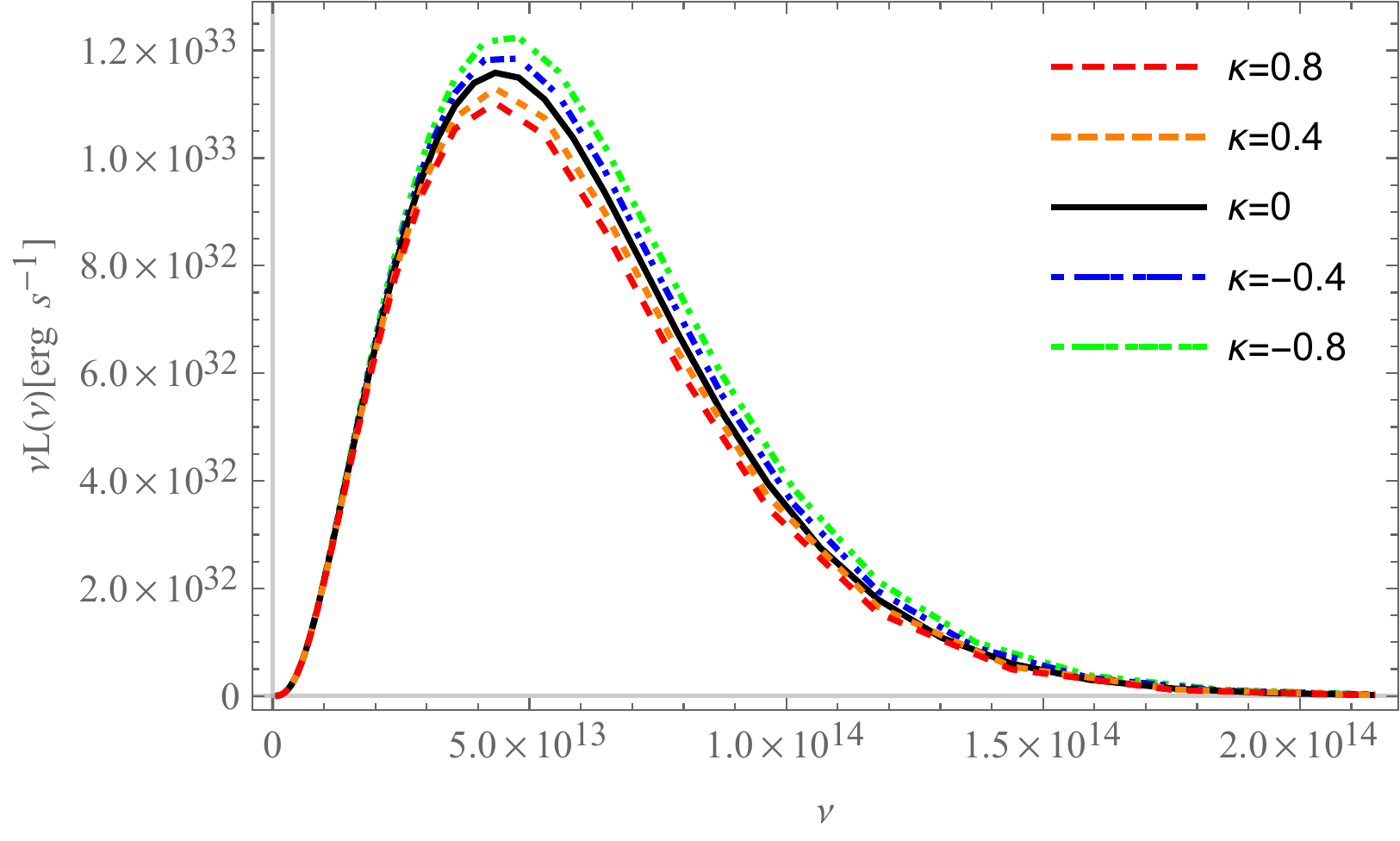}
\includegraphics[width=3.4in]{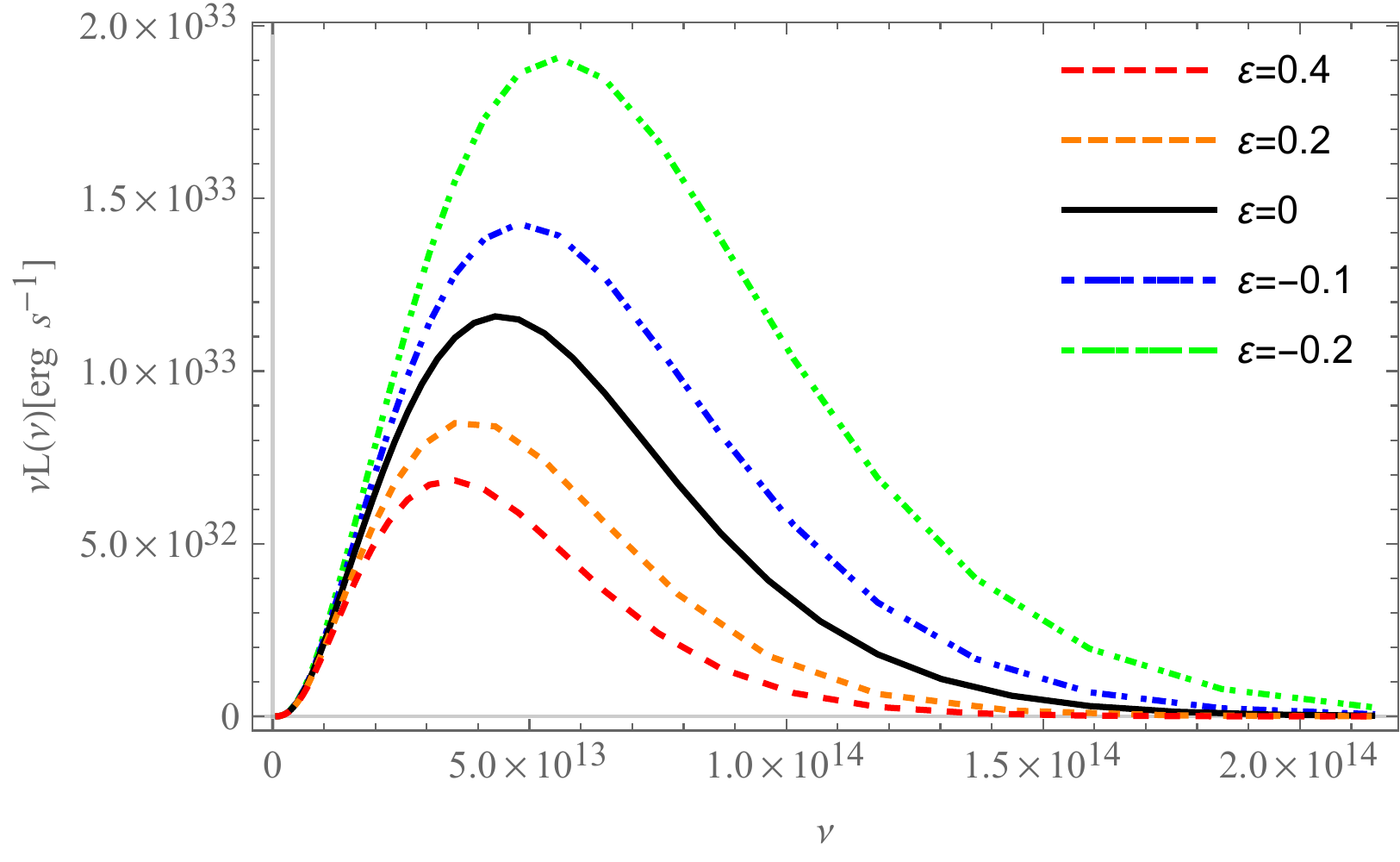}
	\caption{\red{The emission spectrum profile for the thin accretion disk around the Einstein-\AE ther black hole for different values of the \ae{}ther field coupling constant $\kappa$ (left panel) for the first solution and $\varepsilon$ (right panel) for the second one. Here the mass of the black hole and the mass accretion rate are set to be $10^{6} M_{\odot}$ and $10^{-12}M_{\odot}/{\rm yr}$, respectively. And we only considered $Q=0$, the neutral black hole case.}}
	\label{Lumi}
\end{figure*}

\red{To illustrate the effect of the \ae{}ther field term in the emission spectrum, we calculate the radiation spectrum $\nu L(\nu)$ numerically and display its behavior as a function of the observed frequency $\nu$ for different type of solutions with different values of the \ae ther field coupling constant $\kappa$ and $\varepsilon$ in Fig.~\ref{Lumi}. For negative $\kappa$ and $\varepsilon$, it is shown that the decreasing values of $\kappa$ and $\varepsilon$ produce greater maximal amplitude of the disk emission spectrum as compared to the standard Schwarzschild case, while for positive $\kappa$ and $\varepsilon$, it produces a smaller maximal amplitude. From these figures, one also observes that the cut-off frequencies of the emission spectra increases for the negative $\kappa$ and $\varepsilon$) and decreases for positive $\kappa$ and $\varepsilon$, from the value corresponding to the standard Schwarzschild black hole. Also due to the different order of the correction term of the \ae{}ther field, the change of the \ae{}ther field coupling constant of the second type of solution has a much greater influence on the radiation luminosity of the accretion disk than that of the solution of the first type.}

\red{At last, let us consider the accretion efficiency of the Einstein-\AE ther black hole}, which is defined as the ratio of the rate of the radiation of energy of photons escaping from the disk surface to infinity and the rate at which mass-energy is transported to the black hole \cite{Novikov1973, Page1974}. If all the emitted photons can escape to infinity, one can find that the efficiency $\epsilon$ is related to the specific energy of the moving particle in the disk measured at the marginally stable orbit by
\bqn
\eta = 1- \tilde E_{\rm ms}.
\eqn
\begin{figure*}
\includegraphics[width=3.41in]{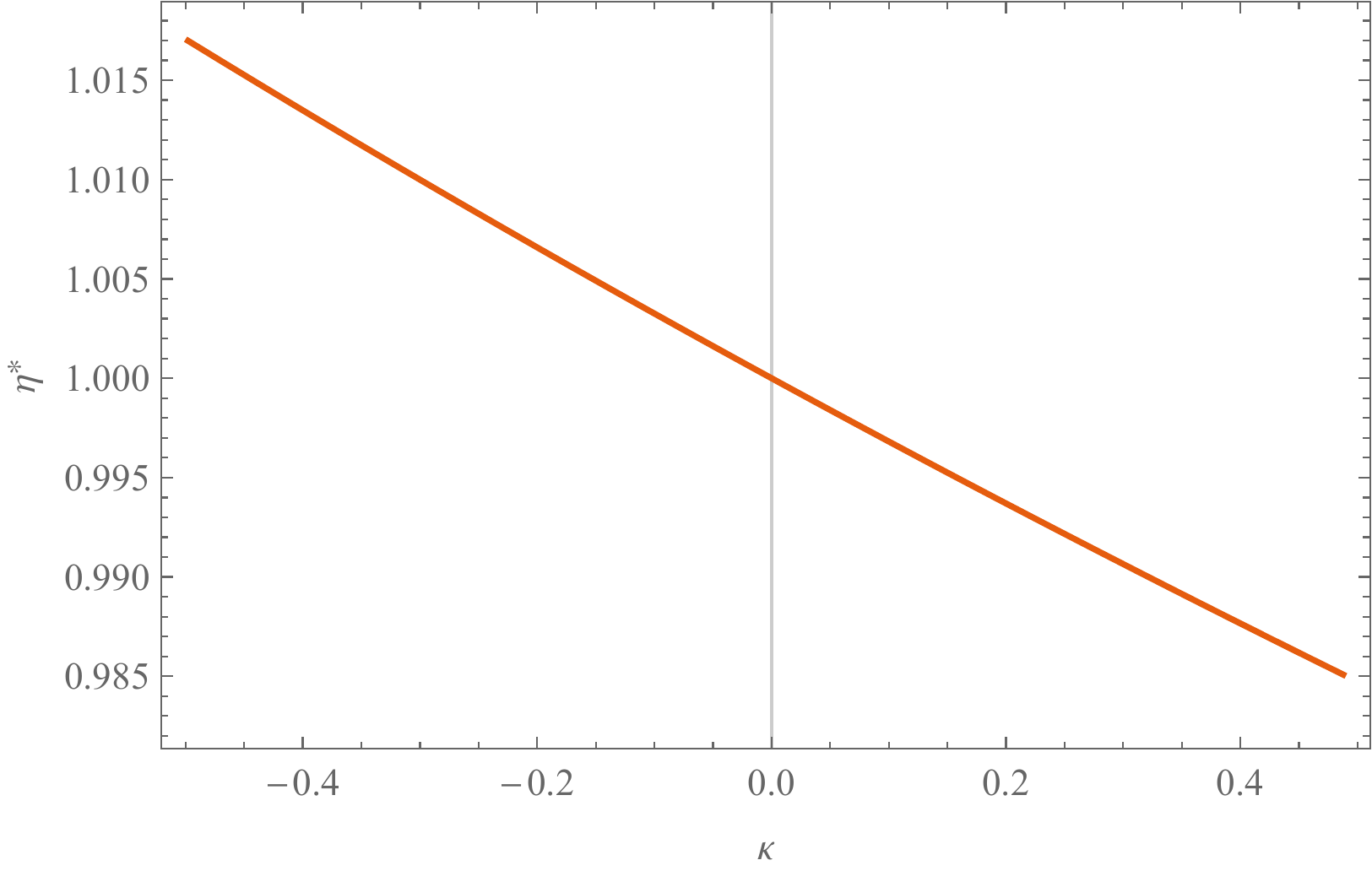}
\includegraphics[width=3.41in]{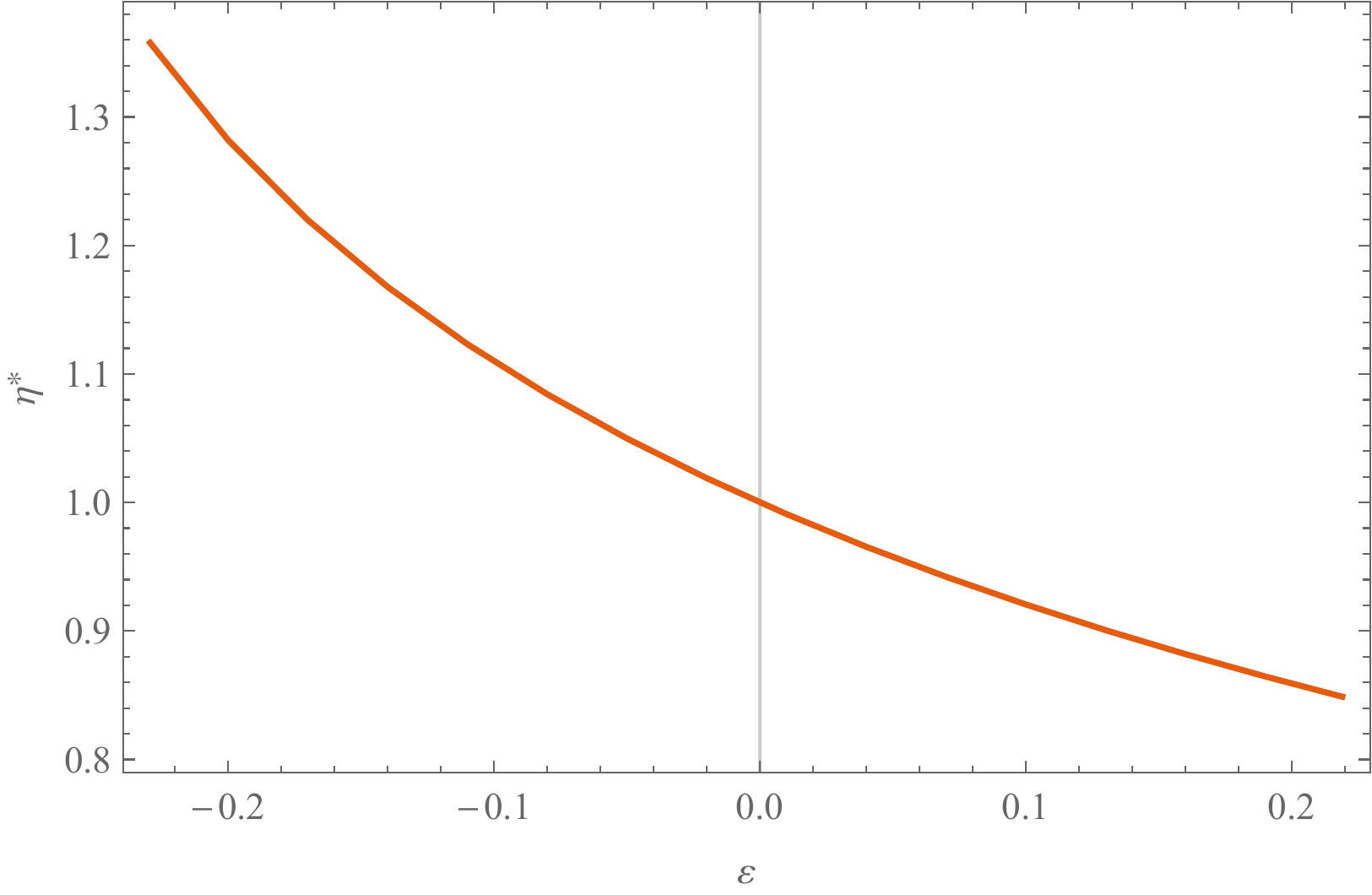}
\caption{\red{The specific accretion efficiency  $\eta^*=\eta/\eta_{Sch}$ of the Einstein-\AE ther black hole as a function of the \ae ther field coupling constant $\kappa$ (left panel) for the first solution and $\varepsilon$ (right panel) for the second one. Here we only considered $Q=0$, the neutral black hole case.}}
%Here the mass of the black hole and the mass accretion rate are set to be $10^{6} M_{\odot}$ and $10^{-12}M_{\odot}/{\rm yr}$ respectively.
\label{eta}
\end{figure*}

\red{The dependence of the accretion efficiency $\eta$ on the \ae ther field coupling constant $\kappa$ and $\varepsilon$ is plotted in Fig.~\ref{eta}. It shows that the accretion efficiency $\eta$ decreases with the increasing values of coupling constant $\kappa$ for the first type of solution and $\varepsilon$ for the second type of solution. This indicates that the accretion of matter in two types of the Einstein-\AE{}ther black hole is more efficient for the negative $\kappa$ and $\varepsilon$ and lesser efficient for the positive $\kappa$ and $\varepsilon$ than that in the Schwarzschild black hole. Therefore, the two types of Einstein-\AE ther black hole with negative \ae ther field coupling constant $\kappa$ and $\varepsilon$ can provide a more efficient engine for transforming the energy of accreting matter into electromagnetic radiation than that with a positive value of $\kappa$ and $\varepsilon$.}

\section{Images of the slowly rotating black holes with accretion disk}
\red{In this section, we investigated the physical model of radiation of the thin accretion disk consisting of particles moving on circular equatorial orbits around the two types of Einstein-\AE ther black holes, and studied its image as seen by a distant observer.} For this purpose, we construct accretion disk images for these black holes with numerical ray-tracing techniques discussed in \cite{Chen_bin}. \red{For the ray-tracing geometry, we define the set of Cartesian coordinates $(X_i, Y_i)$ on the image plane such that the $Y$-axis is along the same fiducial plane and the $X$-axis is perpendicular to it. We then convert the coordinates ($X, Y$) of a photon that reaches the image plane to the coordinates ($r , \theta , \phi$) in the spherical-polar system used for the metric with the relations
\bqn
r&=&\sqrt{D^2+X^2+Y^2},\\
\cos \theta &=& \frac{D \cos\theta + Y \sin\theta}{r}, \\
\tan \phi &=& \frac{X}{D\sin\theta -Y\cos\theta}.
\eqn
The photons that contribute to the image of the compact object are those with 3-momenta
that are perpendicular to the image plane. This orthogonality condition uniquely specifies the
momentum vector of a photon with the above coordinates, according to the relation
\bqn
k^r &\equiv&\frac{dr}{d\lambda'} = \frac{D}{r}, \\
k^\theta &\equiv& \frac{d\theta}{d\lambda'}= \frac{-\cos\theta +\frac{D}{r^2}(D\cos\theta + Y\sin\theta)}{\sqrt{r^2-(D\cos\theta+Y\sin\theta)^2}},\\
k^\phi &\equiv& \frac{d\phi}{d\lambda'} =\frac{-X\sin\theta}{(D\sin\theta-Y\cos\theta )^2+ X^2}. \label{Runge3}
\eqn
With these relations, we can integrate the geodesic equations backward in time from any detection point $(X_0, Y_0, 0)$ in the image plane of the distant observer to the emission point in the disk:
\bqn
\frac{d^2 x^\mu}{d\lambda^2}+\Gamma^\mu_{\nu\rho}\frac{dx^\nu}{d\lambda}\frac{dy^\rho}{d\lambda} = 0
\eqn
where $\lambda$ is an affine parameter. In the numerical algorithm, we integrate geodesic equations using a Runge-Kutta integrator, which is integrated in the ode45 command of the Matlab.}

% \begin{figure}
% \includegraphics[width=2.80in]{rayplot.png}
% \caption{\red{Ray-traced diagram for thin accretion disk model. See the text for more details.}}
% \label{image1}
% \end{figure} 
\begin{figure*}
\includegraphics[width=3.10in]{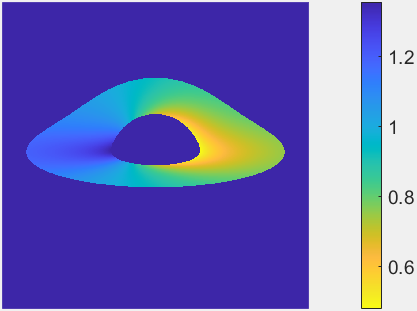}
%~~~~\includegraphics[width=3.11in]{c14_0.001_fig2.png}
~~~~\includegraphics[width=3.07in]{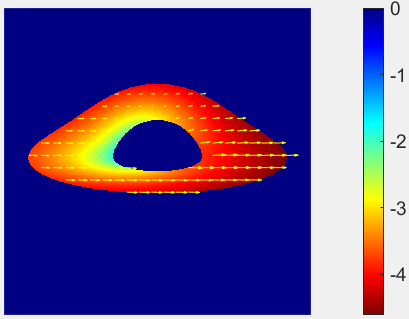}
\caption{Ray-traced redshifted image (left panel) and intensity and polarization profiles (right panel) of a lensed accretion disk around the Schwarzschild black hole, which corresponding to the case of $\kappa=0$ for the first solution or $\epsilon=0$ for the second solution. We present the images for inclination angle $i=75\degree$. See the text for more details.}
\label{image1}
\end{figure*}

\begin{figure*}
\includegraphics[width=3.10in]{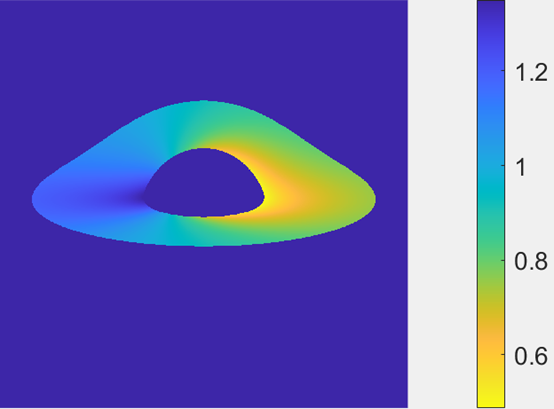}
~~~~~\includegraphics[width=3.10in]{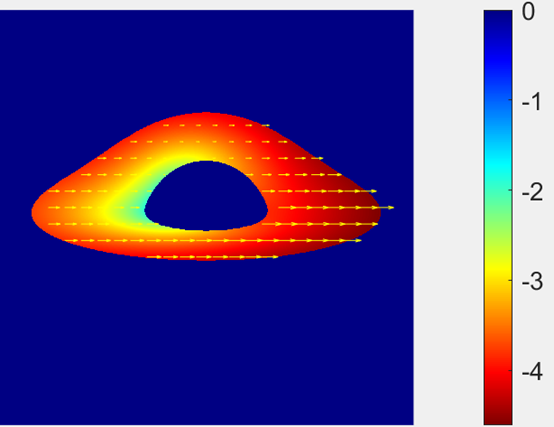}
\caption{Ray-traced redshifted image (left panel) and intensity and polarization profiles (right panel) of a lensed accretion disk around the first type neutral Einstein-\AE ther black hole. We present the images for inclination angle $i=75\degree$ and coupling constant $\kappa=0.6$. See the text for more details.}
\label{image2}
\end{figure*}

\begin{figure*}
\includegraphics[width=3.10in]{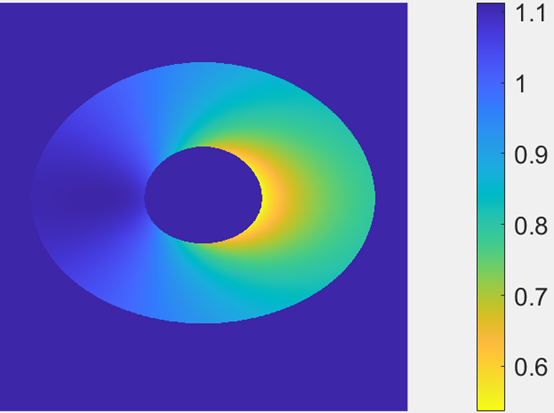}
~~~~\includegraphics[width=3.10in]{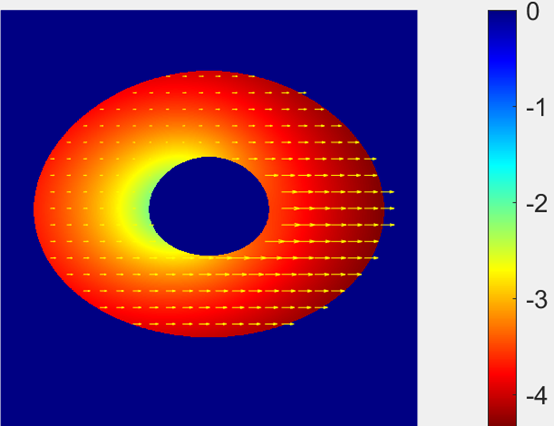}
\caption{Ray-traced redshifted image (left panel) and intensity and polarization profiles (right panel) of a lensed accretion disk around the first type neutral Einstein-\AE ther black hole. We present the images for inclination angle $i=45\degree$ and coupling constant $\kappa=0.6$. See the text for more details. }
\label{image3}
\end{figure*}

\begin{figure*}
\includegraphics[width=3.10in]{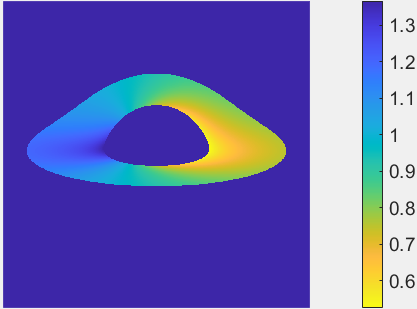}
~~~~~\includegraphics[width=3.07in]{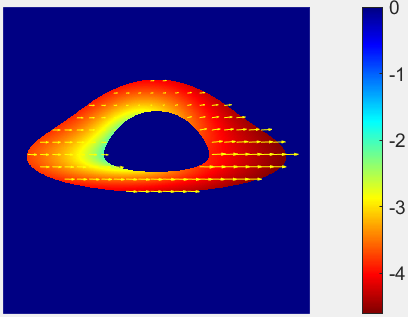}
\caption{Ray-traced redshifted image (left panel) and intensity and polarization profiles (right panel) of a lensed accretion disk around the second type of neutral Einstein-\AE{}ther black hole. We present the images for inclination angle $i=75\degree$ and coupling constant $\epsilon=0.60$. See the text for more details. }
\label{image5}
\end{figure*}

\begin{figure*}
\includegraphics[width=3.08in]{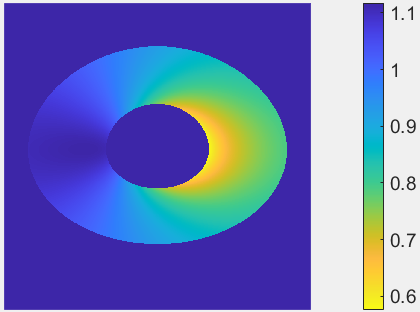}
~~~~\includegraphics[width=3.05in]{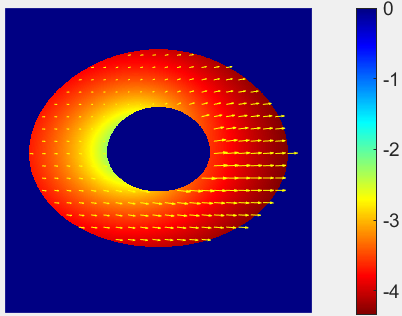}
\caption{Ray-traced redshifted image (left panel) and intensity and polarization profiles (right panel) of a lensed accretion disk around the second type of neutral Einstein-\AE{}ther black hole. We present the images for inclination angle $i=45\degree$ and coupling constant $\epsilon=0.60$. See the text for more details. }
\label{image6}
\end{figure*}

According to the fact that accretion disk radiation emission is observed to follow a power law, we assume the local source intensity of the radiation emitted by the disk is
\bqn
I_\nu(\nu,\mu,r) \propto \frac{1}{r^n}\frac{w(\mu)}{\nu^(\Gamma-1)},
\eqn
where $\nu$ is the photon frequency in the rest frame of the gas, $\mu$ is the cosine between the photon's 4 momentum and the upward disk normal measured by the comoving observer, $w(\mu)$ is the angular-dependence of the intensity profile, $\Gamma$ is a photon index and n represent the radial steepness of the intensity profile.

In addition, considering a well-known fact that a photon's polarization vector is parallel transported along the photon's geodesic and its plane of polarization is rotated as the photon passes the rotating black hole, which is called gravitational Faraday rotation, the radiation from the disk becomes partially polarized. Another factor affecting polarization is the Thomson scattering of photons off free electrons in the dense atmosphere of the disk, the degree of polarization depends on the angel $i$ between the normal to the disk surface and the direction of propagation of the radiation photon, ranging from 0 ($i=0\degree$, photon direction parallel to the normal to the disk) to about $12\%$ ($i=90\degree$, photon direction orthogonal to the normal to the disk). See the references \cite{polarization1, polarization2} for more details.

\red{In Figure \ref{image1} and \ref{image2}, we plot the ray-traced  redshifted  image  of  a  lensed  accretion  disk  (left  panel)  as well as the intensity and polarization  profile (right panel)  of  an accretion disk around the first solution of the Einstein-\AE ther theory observed at inclination angle 75° but for different values $\kappa=0$ and $\kappa=0.6$. The former actually corresponds to the image of the thin accretion disk of the Schwarzschild black hole, which serves as a comparison for us to discuss the effects of the \ae ther field. Figure \ref{image3} is similar to Figure \ref{image2}, but for inclination angle $i = 45\degree$. Figures \ref{image5}, \ref{image6} are the same panel of graphs as Figure \ref{image2}, \ref{image3}, but corresponds to the thin accretion disk in the second Einstein-\AE{}ther spacetime which coupling constant takes the value $\varepsilon=0.6$. For the ray-traced redshifted image, its color bar represents the degree of redshift of light emitted from the thin accretion disk. And for the intensity and polarization profile map, the color bar describes the intensity of radiation on the accretion disk, in addition, the arrows in the figure represent the polarization of the radiation light, and the length and direction of the arrows indicate the size and direction of the polarization, respectively.  }

As we all known, astrophysical black holes, whether stellar mass or supermassive, generally have strong gravitational lensing. It changes both the shape and the intensity profile of the disk significantly. The intensity is strongly concentrated in a small region (in dark red) near the black hole and to the left of the accretion disk (where the source is approaching the observer). The hat-like structure shown in the figures \ref{image1} - \ref{image5} are formed by bending the flat structure of the accretion disk due to the gravitational lensing effect of the black hole.

% As for polarization, the degree of polarization to the left of the black hole is on average smaller than the classical value (∼4.6\% for θ = 75°; Chandrasekhar 1960) because the angle between the photon’s 4 momentum and the disk’s normal measured by the comoving observer is smaller than the inclination angle.

\red{As we can see from the these figures, the ray-traced redshifted images and intensity and polarization profile of the thin accretion disk around both the type I and type II Einstein-\AE ther black holes resemble closely the Schwarzschild case. In spite of we choose an Einstein-\AE ther spacetime with a value of the \ae ther field coupling constant with maximally deviating from the Schwarzschild solution, only slight quantitative differences are present. For the first Einstein-\AE{}ther black hole solution, one can see that it is very little different from the Schwarzschild black hole case. Of course, this is also very easy to understand, for this solution correction to Schwarzschild space-time is on the fourth-order term. So for the space-time structure that is slightly farther away from the black hole, it is very difficult to see how it is different from the space-time of the Schwarzschild black hole.}

\red{For the second type of the Einstein-\AE{}ther solution, the most interesting point is that the magnification of the coupling constant has a significant impact on the disk central shadow area of the accretion disk, which is particularly evident in Figures \ref{image1} and \ref{image5}, the disk central shadow area of the accretion disk increases gradually with the increase of the coupling constant $\varepsilon$. This result is completely consistent with the conclusions discussed by the authors in \cite{Tao_shadow}. The central shadow area of the accretion disk corresponds to the shadow region of the black hole, which changes with the change of the \ae ther field parameters. From Figures \ref{image1} and \ref{image2}, \ref{image1} and \ref{image5}, we can see that the coupling constants $\kappa$ and $\varepsilon$ has a very limited effect on the redshift distribution of radiated light emitted from the accretion disk. In contrast, the inclination angle has a more obvious effect on the shape of both redshifted imege and intensity and polarization profile, as can be seen from figures \ref{image2} and \ref{image3} as well as figures \ref{image5} and \ref{image6}.}

\red{On the other hand, we know that the magnification and direction of radiation polarization can change significantly since the gravitational Faraday rotation. This effect is more significant for source regions closer to the black hole where the observed polarization angle $\chi$ can be positive or negative (depending on the actual source location). Since what we use here is the static, spherically symmetric Einstein-\AE{}ther black hole solution, the \ae ther field has no significant contribution to the gravitational Faraday rotational effect of light around the black hole, resulting in the \ae ther field having no effect on the magnification and direction of polarization. However, as we can see from Figure \ref{image2} and \ref{image3}, \ref{image5} and \ref{image6}, the magnification increases with inclination angles and is more significant for steeper profiles. And for moderate to high inclination angles, the emission is strongly focused in a small region with a low degree of polarization. }

\section{Conclusion and Discussion}

\red{In this paper, we study the physical properties of a thin accretion disk around two types of the black hole solutions in Einstein-\AE{}ther theory. The physical quantities of the thin accretion disk, such as the energy flux, temperature profile, electromagnetic emission spectrum profiles, and the accretion efficiency have been analyzed in detail for the two types of the black hole solutions. The effects of the \ae ther field coupling constant $\kappa$ for the first solution and $\varepsilon$ for the second solution in Einstein-\AE{}ther theory on these physical quantities have been explicitly obtained. For the first type of the black hole solution, it is shown that with the increases of the parameter $\kappa$, energy flux, temperature distribution, and electromagnetic spectrum of the disk all decreases. For the second type of solution, it is shown that with the increases of the parameter $\varepsilon$, energy flux, temperature distribution, and electromagnetic spectrum of the disk all decreases as the first black hole solution case, however, the variation of these physical quantities with the \ae{}ther field coupling constant is much larger than that of the first type of black hole solution. The main reason for this kind of behaviors is that the correction of the first type of black hole solution to space-time is on the fourth-order term of $M/r$, and the impact on the physical properties of the surface of the accretion disk is much smaller than that of the second type of black hole solution corrected by the \ae{}ther field on the second-order term.}  %for positive $c_{13}$ or negative $c_{14}$, the effect of the \ae ther field coupling constant $c_{13}$ and $c_{14}$ increases the radius of the marginally stable orbit so that the lower limit of the integral in (\ref{energyflux}) becomes smaller, while for negative $c_{13}$ or positive $c_{14}$ the radius of the marginally stable orbit increases so that the lower limit becomes smaller.} 

\red{In addition, we also show that the accretion efficiency decreases as the growth of the coupling constant $\kappa$ for the first solution and $\varepsilon$ for the second solution. Our results indicate that the thin accretion disk around the black hole in Einstein-\AE ther theory is hotter, more luminosity, and more efficient than that around a Schwarzschild black hole with the same mass for the first type of Einstein-\AE{}ther black hole with a negative $\kappa$ and the second type of solution with a negative $\varepsilon$, while it is cooler, less luminosity, and less efficient for a positive $\kappa$ and a positive $\varepsilon$. Similarly, since the different order of the correction term of the \ae{}ther field, the variation range and amplitude of the accretion rate of the second type of black hole with the \ae{}ther field coupling constant is also much larger than that of the first type of black hole. }

\red{Then we made the image simulation for the thin accretion disk around the two types of Einstein-\AE{}ther black holes. In this part we observe no qualitative distinction in the appearance for both the ray-traced redshifted and intensity and polarization profile of the thin accretion disk around the two types of the Einstein-\AE{}ther black holes and the Schwarzschild black hole. Only small quantitative differences are present in the disk size and the maximum of the radiation flux. We found that the \ae ther field has a certain impact on the disk surface area of the accretion disk, the disk surface area of the accretion disk increases gradually with the increase of the coupling constant $\varepsilon$ for the second type of the solution. For the first type of black hole, it is difficult to see any obvious changes from the image simulation, although our parameter values are as large as possible.} \red{Moreover, for the intensity and polarization profile image, we find that the observed polarized intensity in the bright region is stronger than that in the darker region. It is also noted that the effect of \ae{}ther field on the observed polarized vector is weak in general and the stronger effect of \ae{}ther appears in the bright region close to black hole in the image plane. These features in the polarized image could help us to understand black hole shadow, thin accretion disk and the coupling between photon and \ae{}ther field. We describe in detail the physical mechanism for the formation of the accretion disk image showing that the phenomenon is expected also for other spacetimes possessing the same properties of the photon dynamics.}  

\red{With the above main results, it is of interest to constrain the black hole parameters, including the angular momentum and the \ae ther field coupling constant $\kappa$ and $\varepsilon$, by using the observation spectra of the X-ray binaries. For example, by using the continuum-fitting method \cite{Cosimo1} or quasi-periodic oscillations \cite{Cosimo3}, one is able to measure the angular momentum of the stellar-mass black holes and constrain the deviations from Kerr metric by using X-ray data from black hole binaries.  On the other hand, we aim at investigating the observable features of the two types of the Einstein-\AE{}ther black holes, which are not present for Schwarzschild black holes and could serve as an experimental test for distinguishing the two types of black holes.}

\section*{ACKNOWLEDGEMENTS}

This work is supported in part by the Zhejiang Provincial Natural Science Foundation of China under Grant No. LR21A050001 and LY20A050002, the National Key Research and Development Program of China Grant No.2020YFC2201503, the National Natural Science Foundation of China under Grant No. 11675143, and the Fundamental Research Funds for the Provincial Universities of Zhejiang in China under Grant No. RF-A2019015.
%This work is supported by National Natural Science Foundation of China with the Grants No.11675143, the Zhejiang  Provincial  Natural Science Foundation of China under Grant No. LY20A050002, and the Fundamental Research Funds for the Provincial Universities of Zhejiang in China under Grants No. RF-A2019015.

\end{document}